\documentclass[aps,prl,reprint,amssymb,showpacs,superscriptaddress,notitlepage,onecolumn,floatfix]{revtex4-2}
\usepackage{bm}
\usepackage{graphicx}
\usepackage[dvipsnames]{xcolor}
\usepackage{hyperref}
\usepackage{dcolumn}
\usepackage{braket}
\usepackage{natbib}
\usepackage{amsmath}
\usepackage{color}
\usepackage{mathtools,amssymb}
\usepackage{soul}
\usepackage{amssymb}
\usepackage{esint}
\usepackage{mathtools}
\usepackage{physics}
\usepackage{makecell}
\usepackage{multirow}
\usepackage{array}
\usepackage{geometry}
\usepackage{url}
\usepackage{adjustbox}
\usepackage{rotating}
\usepackage{listings}
\usepackage{threeparttable,booktabs}
\usepackage{indentfirst}
\usepackage{fontawesome}
\usepackage{comment}
\usepackage{cleveref}
\definecolor{beaver}{rgb}{0.62, 0.51, 0.44}
\definecolor{bole}{rgb}{0.47, 0.27, 0.23}
\definecolor{brown(traditional)}{rgb}{0.59, 0.29, 0.0}
\definecolor{burntorange}{rgb}{0.8, 0.33, 0.0}
\definecolor{RedBlue}{rgb}{0.8, 0, 0.8}

\makeatletter
\newenvironment{supplement}{%
  \setcounter{secnumdepth}{3}%
  \setcounter{equation}{0}

  \setlength{\parindent}{1.5em}
  \setlength{\parskip}{0pt}

  \setcounter{section}{0}%
  \def\@seccntformat##1{\csname the##1\endcsname:\ }%

  \setcounter{figure}{0}%
  \setcounter{table}{0}%
}{%
  \def\@seccntformat##1{\csname the##1\endcsname.\ }%
}
\makeatother

\begin{document}
\title{Single-Shot Realization of 10000-Mode Octave-Spanning Artificial Gauge Fields}

\author{Lida Xu$^{\dagger}$}\email{lidaxu66@umd.edu}
\affiliation{Joint Quantum Institute, University of Maryland and National Institute of Standards and Technology,
College Park, MD 20742, USA}

\author{Apurva Padhye}
\altaffiliation{Equally contributing authors}
\affiliation{Joint Quantum Institute, University of Maryland and National Institute of Standards and Technology,
College Park, MD 20742, USA}

\author{Supratik Sarkar}
\affiliation{Joint Quantum Institute, University of Maryland and National Institute of Standards and Technology,
College Park, MD 20742, USA}

\author{Alireza Parhizkar}
\affiliation{Joint Quantum Institute, University of Maryland and National Institute of Standards and Technology,
College Park, MD 20742, USA}

\author{Christopher J. Flower}
\affiliation{Joint Quantum Institute, University of Maryland and National Institute of Standards and Technology,
College Park, MD 20742, USA}

\author{Gr\'egory Moille}
\affiliation{Joint Quantum Institute, University of Maryland and National Institute of Standards and Technology,
College Park, MD 20742, USA}

\author{Kartik Srinivasan}
\affiliation{Joint Quantum Institute, University of Maryland and National Institute of Standards and Technology,
College Park, MD 20742, USA}

\author{Mohammad Hafezi}\email{hafezi@umd.edu}
\affiliation{Joint Quantum Institute, University of Maryland and National Institute of Standards and Technology,
College Park, MD 20742, USA}

\author{Mahmoud Jalali Mehrabad$^\dagger$}\email{mjalalim@umd.edu}
\affiliation{Joint Quantum Institute, University of Maryland and National Institute of Standards and Technology,
College Park, MD 20742, USA}

\begin{abstract}
Artificial gauge fields (AGFs)~\cite{dalibard2011colloquium,gross2017quantum,ding2025progress,aidelsburger2018artificial} enable photons~\cite{wang2009observation,fang2012realizing,rechtsman2013photonic,roushan2017chiral,dutt2020single,rosen2024synthetic,hafezi2011robust,mittal2019photonica,mittal2019photonicq,mehrabad2025quantum,suh2024photonic} and other bosons~\cite{dalibard2011colloquium,susstrunk2015observation,fang2017generalized,galitski2019artificial,klembt2018exciton,gardin2024engineering} to emulate fermionic phenomena such as chiral edge transport and quantum Hall phases~\cite{hafezi2007fractional,clark2020observation,deng2022observing,wang2024realization}; however, existing theories and realizations remain confined to narrow bandwidths under single-mode approximation. We introduce a general theoretical framework for ultra-broadband, multi-modal dispersion-corrected AGFs in both linear and nonlinear regimes. Using integrated photonics, we realize over 100 distinct AGFs hosting more than 10,000 modes across nearly an optical octave — the first frequency-comb realization of the integer quantum Hall model for photons. Leveraging Kerr nonlinearity, we achieve single-shot AGF control beyond waveguide dispersion, robust to wafer-scale fabrication variations. Our results establish a new regime of ultra-broadband multimodal AGFs, opening pathways to exotic dispersion-corrected AGF dynamics and simulations, as well as volume-manufacturable device functionalities such as waveguide-dispersion-resilient photonic circuits, and AGF-enabled programmable nonlinear and quantum optics and optoelectrics.
\end{abstract}
\maketitle
\section*{Introduction}
Artificial  (or
synthetic) gauge fields (AGFs)~\cite{dalibard2011colloquium,song2025artificial,aidelsburger2018artificial} have recently emerged as powerful ``substitutes" for real electromagnetic fields for charge-neutral particles such as photons, which are normally unaffected by real electromagnetic fields. AGFs enabled the emulation of exotic phenomena such as quantum Hall phases, orbital magnetism, Landau quantization, and topological insulators. In recent years, vast classes of scalar, vector, real, and complex AGFs have been realized in real or synthetic~\cite{celi2014synthetic,ozawa2019topological} dimensions by engineering the tunneling parameters, enabling neutral bosons to accumulate controlled Aharonov–Bohm phases and form topological Bloch bands similar to those in solid state systems. These ideas have been widely explored with ultracold atoms~\cite{dalibard2011colloquium,galitski2019artificial,dalibard2011colloquium,lin2009synthetic,gross2017quantum}, circuit-QED~\cite{roushan2017chiral,rosen2024synthetic,qian2025programmable}, polaritons~\cite{klembt2018exciton,whittaker2021optical}, magnons~\cite{gardin2024engineering,li2025topological}, phonons~\cite{susstrunk2015observation}, opto-mechanics~\cite{fang2017generalized,chen2021synthetic,slim2025programmable}, non-Abelian~\cite{yang2024non,cheng2025non,dong2025sl,lai2025non}, interacting~\cite{hafezi2007fractional,clark2020observation,deng2022observing,wang2024realization} and single-particle photonic lattices~\cite{hafezi2011robust,mittal2019photonica,mittal2019photonicq,mehrabad2025quantum}. 

Among leading platforms for the exploration of AGFs, integrated photonics is particularly attractive, due to room-temperature operation, ultra-broad bandwidth, many available spatial-spectral-temporal degrees of freedom, strong nonlinearity, and compact and efficient volume-manufacturability via CMOS foundries. In photonics, AGFs can be created using magneto-optic materials~\cite{wang2009observation}, symmetry-broken photonic crystals~\cite{barczyk2024observation,barsukova2024direct,jalali2024strain,jalali2020semiconductor,jalali2020chiral}, dynamic modulation~\cite{rechtsman2013photonic}, non-Hermitian synthetic space~\cite{wang2021generating,wang2021topological}, and ring resonator lattices with engineered hopping phases~\cite{hafezi2011robust,mittal2019photonica,mittal2019photonicq,mehrabad2025quantum}. In addition to linear photonic applications such as reconfigurable steering~\cite{zhao2019non}, dispersionless coupling~\cite{song2022dispersionless}, programmable lattices~\cite{dai2024programmable} and switches~\cite{feng2025non}, AGFs enable a wide range of applications in generation~\cite{mittal2018topological} and manipulation~\cite{blanco2018topological,tambasco2018quantum,mittal2021tunable} of quantum light as well as robust lasers~\cite{bandres2018topological,amelio2020theory}, frequency combs~\cite{mittal2021topological,flower2024observation,mehrabad2025quantum}, synchronizers~\cite{xu2025chip}, and harmonic generators~\cite{mehrabad2025multi}. However, nearly all existing theories and realizations remain confined to a single-mode or narrow-band regime because the gauge phase and coupling parameters are strongly frequency dependent. Material dispersion and frequency-sensitive evanescent coupling distort the effective tight-binding Hamiltonian away from its nominal gauge-defined form, preserving a clean topological band structure only over a narrow bandwidth. Theoretical understanding and realization of well-controlled AGFs across many modes and over a broad optical bandwidth, therefore, remains a central and unresolved challenge.
\begin{figure*}[t]
\centering
\includegraphics[width=0.99\textwidth]{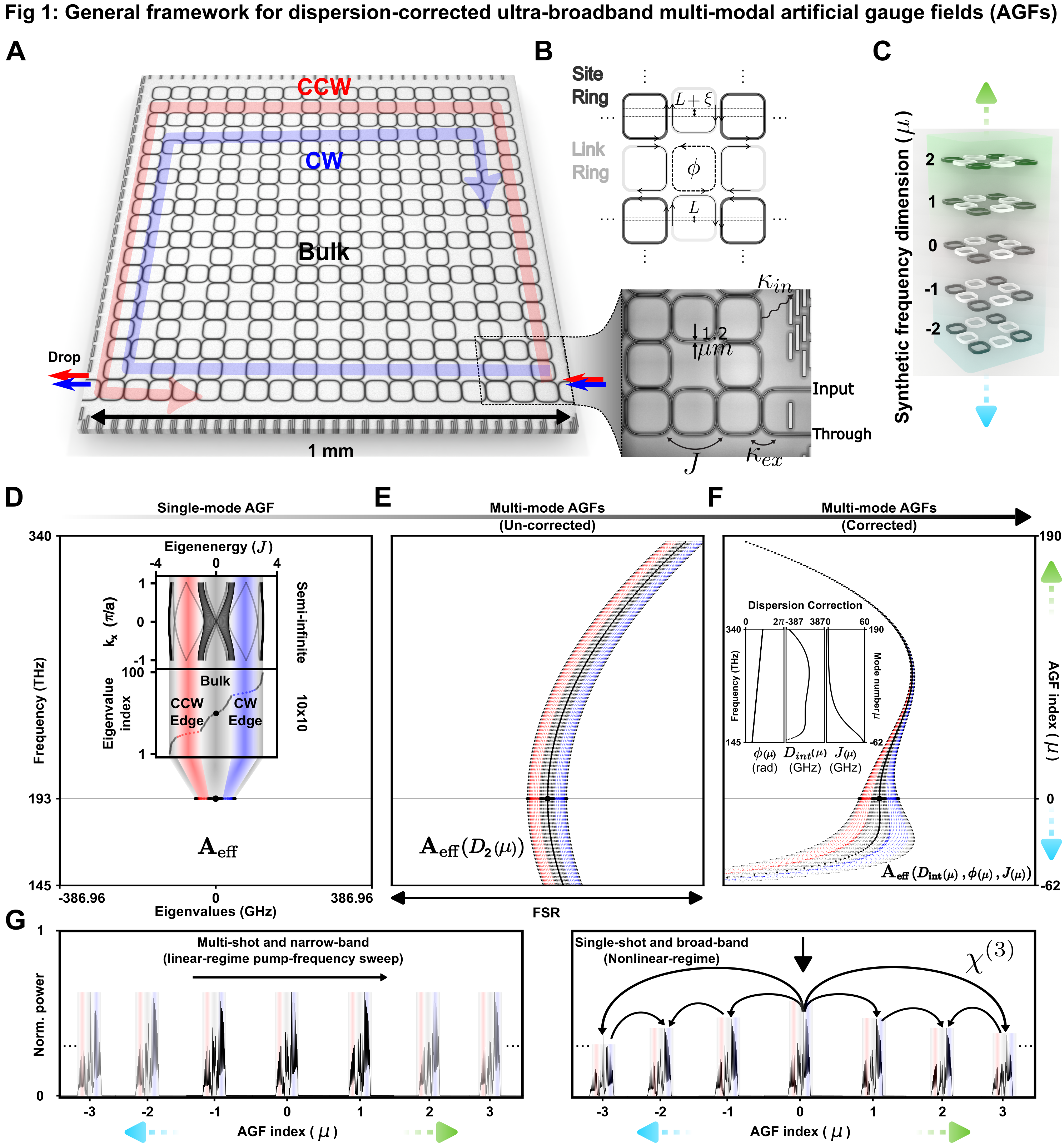}
\caption{\textbf{(A)} High-resolution optical image of the 10~$\times$~10 integer quantum Hall (IQH)
SiN photonic square lattice comprised of a ring resonator array in an add-drop configuration. Two spectrally resolved chiral edge states, characteristics of the IQH model, travel in clockwise/counter-clockwise (CW/CCW)
directions at the lattice boundaries, and are marked in blue and red, respectively. \textbf{(B)} A plaquette of the IQH lattice is
shown both schematically (up) and experimentally (bottom), with the lattice parameters highlighted. Photons completing a round-trip in the plaquette acquire a phase $\phi$. \textbf{(C)} Schematic of multi-modal AGFs in the synthetic frequency dimension indexed by the single-ring mode index $\mu$. \textbf{(D-F)} Eigenvalues of the tight-binding Hamiltonians describing the IQH lattice for cases of single-mode, uncorrected multi-modal, and corrected multi-modal AGFs, respectively. The dispersion-correction plots, shown in the inset of (F), are calculated using broadband 3D FDTD simulations of the SiN waveguide mode. \textbf{(G)} Multi-shot (left) and single-shot (right) realizations of ultra-broadband, multi-modal AGFs. In the multi-shot approach, each realization probes only a narrow pump-frequency window set by the sweep-laser bandwidth, leaving large portions of the AGF spectrum unaddressed (faded regions). The single-shot realization simultaneously accesses the full AGF bandwidth.}
\label{Fig:intro}
\end{figure*}

In this work, we overcome this long-standing barrier by first introducing a minimal theoretical framework for ultra-broadband, multi-modal dispersion-correction that remains valid in both linear response and strongly nonlinear regimes. This framework takes the form of a dispersive tight-binding model in which the on-site potentials, coupling strengths, and hopping phases become frequency-dependent quantities extracted from broadband 3D FDTD simulations of waveguides, enabling accurate modeling of gauge fields across a broad bandwidth. Guided by this theory, we experimentally realize an integer quantum Hall (IQH) lattice of integrated coupled microresonators that supports two chiral edge bands separated by bulk bands. In the linear regime, we realize a small subset of the dispersion-corrected AGFs using multi-shot (pump-frequency-sweeping) transport. Next, we drive the system into a regime where a single-shot pump generates a near octave-spanning IQH frequency comb whose teeth realize more than 100 frequency-dependent AGFs (10000 photonic modes). The dispersion-corrected model quantitatively accounts for both the linear transport spectra and the spatio-spectral profiles of the nonlinear IQH combs realized by a single-shot pump. We then design and realize dispersion-corrected IQH combs in multiple regimes dominantly controlled by the AGF parameters, including combs with highly desirable linearly-dispersive spectral envelopes with remarkably robust bandwidth and spectra across large wafer-scale device variation.
%
\section{Concept}
Fig.~\ref{Fig:intro} introduces the general dispersion-correction framework for ultra-broadband multi-modal AGFs, realized in integrated photonics. The device is a two-dimensional (2D) array of evanescently coupled SiN microring resonators fabricated using commercially available silicon nitride (SiN)~\cite{flower2024observation,xu2025chip,mehrabad2025multi}. A total of 100 site rings, coupled by link rings that are spatially shifted to introduce synthetic magnetic fields, realize a 10~$\times$~10 square lattice that simulates the IQH model for photons~\cite{hafezi2011robust,hafezi2013imaging}, as shown in Figs.~\ref{Fig:intro}A-B. Below, we highlight key features central to the theoretical modeling as well as the single-shot realization of dispersion-corrected ultra-broadband multi-modal AGFs; details of the lattice design are presented in the SI sections~\ref{sm:setup} and~\ref{sm:AQHIQH}.
\subsection{Conventional narrow-band single-mode AGF}
Fig.~\ref{Fig:intro}C-F compares conventional dispersion-uncorrected and dispersion-corrected AGFs. Panel C schematizes the synthetic frequency dimensions formed by the site-ring longitudinal modes $\mu$ that belong to the same transverse-electric (TE) mode family of the SiN waveguide, centered at the telecommunications band. For a single-ring mode $\mu=0$ near 193 THz and a designed effective Landau gauge $\mathbf{A}_{\mathrm{eff}}$, the conventional single-mode tight-binding Hamiltonian description of the IQH lattice~\cite{hafezi2011robust} reads:
\begin{equation}
\begin{aligned}
H_{\mathrm{IQH}}
= \sum_{\substack{x,y}}\omega_0 \hat{a}^{\dagger}_{x,y}\,\hat{a}_{x,y}-\, J \sum_{\substack{x,y}}
\Big( \hat{a}^{\dagger}_{x+1,y}\,\hat{a}_{x,y}\,
e^{-i\,y\,\phi} \;+\;
\hat{a}^{\dagger}_{x,y+1}\,\hat{a}_{x,y} \;+\; \mathrm{h.c.} \Big).
\end{aligned}
\label{eqn:single_mode_tb_H_iqh_main}
\end{equation}
Here $\omega_0$ is the on-site potential, $\hat{a}^{\dagger}_{x,y}$ ($\hat{a}_{x,y}$) is the photon creation (annihilation) operator at the site ring labeled by a Cartesian coordinate $(x,y)$, and $\phi$ is the hopping phase accumulated by photons through each plaquette, defined by the synthetic gauge field $\mathbf{A}_{\mathrm{eff}}=(y\phi,0,0)$ (Fig.~\ref{Fig:intro}B). The effective coupling strength between nearest-neighbor site rings is denoted by $J$, while the intrinsic loss rate at each site ring and the extrinsic coupling rate between the input-output waveguides and the resonators are denoted by $\kappa_{in}$ and $\kappa_{ex}$, respectively. The lattice is designed such that for $\mu=0$, the on-site potential $\omega_0=0$ and the hopping phase $\phi=\pi/2$. Fig.~\ref{Fig:intro}D plots the eigenvalues of Hamiltonian~\ref{eqn:single_mode_tb_H_iqh_main} that correspond to the super-modes collectively formed by the site rings, manifesting as two chiral edge bands separated by a bulk band, which is a characteristic signature of the IQH model. The number of super-modes is determined by the lattice size, which here is 100. In particular, there are 10 edge modes indexed by $\sigma$ in each edge band.

Conventionally, due to the typically narrow topological band ($\approx 100$ GHz) for a single mode $\mu$, the AGF parameters $\omega_0$, $J$, and $\phi$ are treated as largely frequency-independent. This simplifying assumption is routinely employed in the single-mode approximation modelings used in previous studies of AGFs in microring resonators~\cite{hafezi2011robust,hafezi2013imaging,dai2024programmable,ma2024anisotropic,li2025tight}. Previous approaches~\cite{mittal2021topological,huang2024hyperbolic,hashemi2025reconfigurable,hashemi2024floquet,flower2024observation,xu2020photonic,mehrabad2025multi} for extending AGFs from single-mode to a multi-mode Hamiltonian description are typically to use copies of the Hamiltonian~\ref{eqn:single_mode_tb_H_iqh_main}, considering only the effects of second-order waveguide dispersion $D_2$ on the on-site potential $\omega_0$, as shown in Fig.~\ref{Fig:intro}E. Those methods still assumes that $J$ and $\phi$ are frequency-independent values across multiple modes $\mu$ separated by free spectral ranges (FSRs) of $\approx$~773.92~GHz, which becomes increasingly invalid in ultra-broadband multi-modal regimes. 
\subsection{General dispersion-correction for ultra-broadband multi-modal AGFs}
We address the bottlenecks of conventional AGFs by presenting a minimal theoretical framework for constructing a dispersive tight-binding Hamiltonian that accurately accounts for the frequency-dependent AGF parameters $J(\mu), D_{\mathrm{int}}(\mu)$ and $\phi(\mu)$~\cite{hafezi2013imaging}: 
\begin{equation}
\begin{aligned}
H_{\mathrm{IQH}}(\mu)
&= \sum_{\substack{x,y}} \,\{\omega_{\mu} -J(\mu)\cot[\zeta\phi(\mu)]\}\;
\hat{a}^{\dagger}_{x,y}(\mu)\,\hat{a}_{x,y}(\mu) \\
&\quad -\, J(\mu)/\sin[\zeta\phi(\mu)] \sum_{\substack{x,y}}
\Big( \hat{a}^{\dagger}_{x+1,y}(\mu)\,\hat{a}_{x,y}(\mu)\,
e^{-i\,y\,\phi(\mu)} \;+\;
\hat{a}^{\dagger}_{x,y+1}(\mu)\,\hat{a}_{x,y}(\mu)\;+\; \mathrm{h.c.} \Big),
\end{aligned}
\label{eqn:dispersive_tight_binding_main}
\end{equation}
\noindent where the resonant frequency of mode $\mu$ is given by $\omega_{\mu}=\omega_0+D_1\mu+D_{\mathrm{int}}(\mu)$. Here, $D_1$ is the FSR and $D_{\mathrm{int}}(\mu)$ is the integrated dispersion of a site ring. Similar to a single ring, we define the integrated dispersion of a lattice as the spectral deviation of its super-modes from an equidistant frequency grid separated by the site ring's FSR, and centered at the pump mode frequency. With this definition, the integrated dispersion of the IQH lattice is calculated by solving the eigenvalues of $H_{\rm{IQH}}(\mu)-\omega_0-D_1\mu$ for each mode $\mu$, as shown in Fig.~\ref{Fig:intro}F. $\zeta$ is a design parameter~\footnote{the ratio between the relative size and shift of the link ring. See SI section~\ref{sm:dtb} for details.} which here is set to one, and $J(\mu),D_{\mathrm{int}}(\mu)$ and $\phi(\mu)$ are obtained from 3D FDTD simulations of waveguides, whose details, together with the derivation of the Hamiltonian~\ref{eqn:dispersive_tight_binding_main}, are provided in the SI sections~\ref{sm:dtb} and~\ref{sm:2ddispersion}. Importantly, the dispersion-corrected AGFs' landscape in Fig.~\ref{Fig:intro}F is markedly different than the conventional counterpart (Fig.~\ref{Fig:intro}E), with an increasing deviation for AGFs further away from the central mode.  
%
\subsection{Narrow-band multi-shot versus ultra-broadband single-shot realization}
Fig.~\ref{Fig:intro}G introduces two schemes to realize the dispersion-corrected multi-modal AGFs. In the multi-shot scheme (left panel), a low-power tunable continuous-wave laser is swept across several resonances accessible within the pump's tuning range, within which a subset of AGFs are realized via linear drop port transmission. This scheme can be modeled using the input/output formalism~\cite{hafezi2014measuring}.

The above multi-shot approach is slow, spectrally limited by the pump's frequency tuning range, and offers no control over the AGFs' properties. Nonlinear effects such as the Kerr effect can provide a contrasting single-shot and massively broadband approach for the realization as well as the control of the AGFs. In the single-shot scheme, a single-wavelength pump above the optical parametric oscillation (OPO) threshold can generate a Kerr frequency comb via cascaded four-wave mixing, enabling the simultaneous realization of ultra-broadband multi-modal AGFs in the nonlinear regime. These AGFs are experimentally detected via optical spectrum analyzer (OSA) measurements of the drop port comb as well as frequency-selective imaging cameras for spatial profiling. We model this scheme by combining the dispersive tight-binding Hamiltonian~\ref{eqn:dispersive_tight_binding_main} with the Lugiato-Lefever equation (LLE)~\cite{chembo2013spatiotemporal,mittal2021topological}:
\begin{equation}
\begin{aligned}
\frac{d a_{x,y}(\mu)}{d t}
&= \,i\delta\,a_{x,y}(\mu)
   \;+i\Big<[\mathcal{H}_\mathrm{IQH}(\mu)-\omega_0-D_1\mu,\hat{a}_{x,y}(\mu)]\Big> \\
&\quad +\, i\gamma\,\mathcal{FT}\mathbf{\{}
     \bigl|E_{x,y}(\theta)\bigr|^{2} E_{x,y}(\theta)\mathbf{\}}\,  \\
&\quad -\,\bigl(\kappa_{\mathrm{ex,\mu}}\,\delta_{\mathrm{IO},(x,y)} + \kappa_{\mathrm{in}}\bigr)\,a_{x,y}(\mu)
     \;+\; \delta_{\mathrm{IO},(x,y)}\delta_{\mu,0}\,\sqrt{2\kappa_{\mathrm{ex,\mu}}}\mathcal{E}\,.
\end{aligned}
\label{eqn:cmt_iqh}
\end{equation}
\noindent where $a_{x,y}(\mu)\equiv\Big<\hat{a}_{x,y}(\mu)\Big>$ is the electric field, $ \delta=\omega_p-\omega_0$ is the pump detuning with respect to the site ring resonance, $\gamma$ is the nonlinear coefficient, $\mathcal{E}$ is the pump amplitude, $\delta_{\mathrm{IO},(x,y)}$ is zero for most rings and one for the input/output rings, and $\mathcal{FT}$ denotes the Fourier Transform. The intra-cavity electric field for a site ring with lattice coordinate ($x,y$) is $E_{x,y}(\theta)$, where $\theta$ is the azimuthal coordinate inside the ring. $E_{x,y}(\theta)$ is the Fourier transform of the field amplitude in the mode-number basis $a_{x,y}(\mu)$. More details about equation~\ref{eqn:cmt_iqh} are presented in the SI section~\ref{sm:lle}.
\begin{figure*}[t]
\centering
\includegraphics[width=0.99\textwidth]{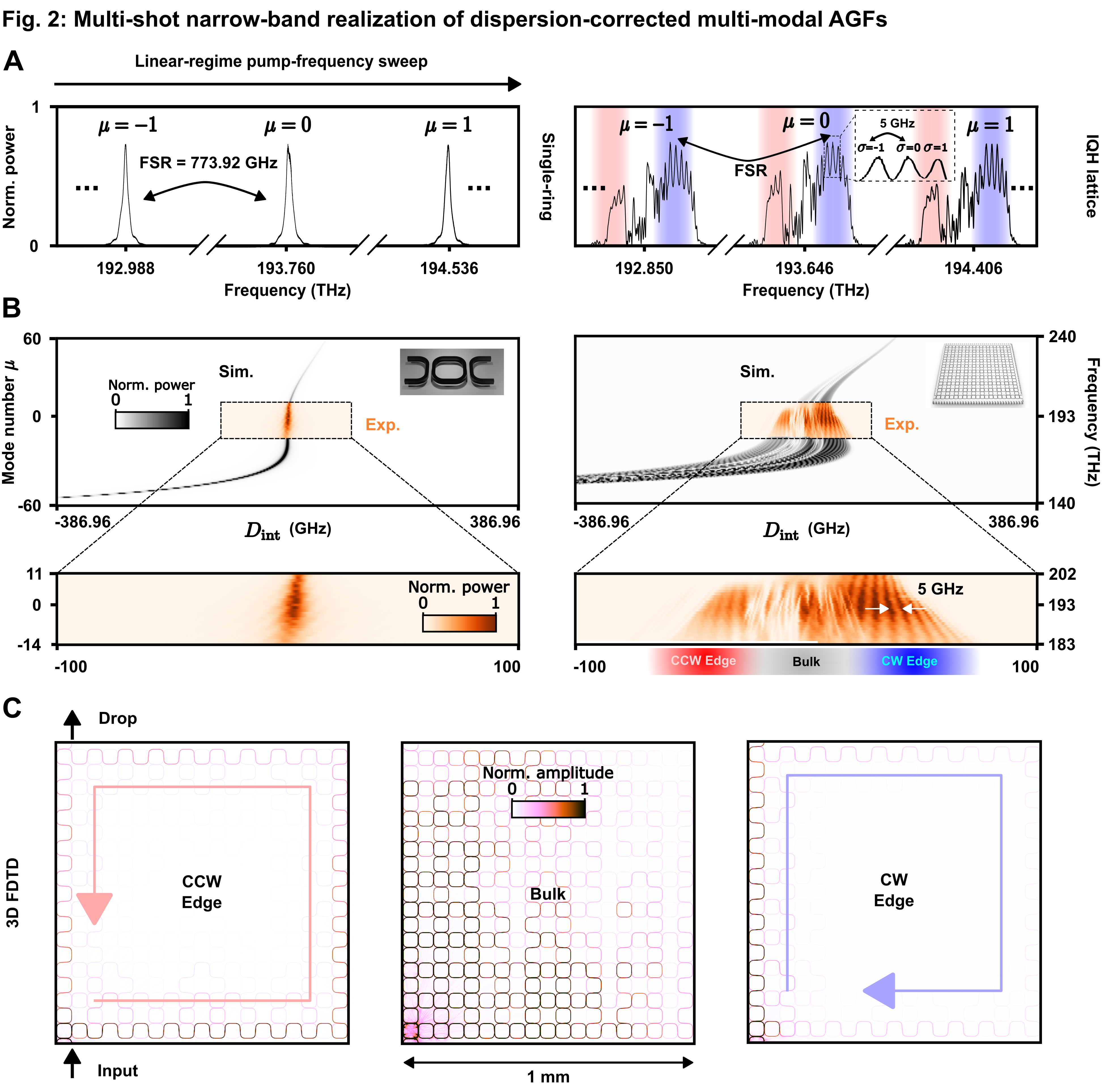}
\caption{\textbf{(A)} Drop port linear transmission spectra of a SiN (left) single ring-resonator and (right) 10~$\times$~10 IQH lattice, measured by sweeping the pump frequency. Each single-ring resonance is indexed by a mode number $\mu$, spectrally separated by the FSR = 773.92 GHz. For the IQH lattice, each single-ring resonance is replaced by the IQH spectra corresponding to its topological band structure, i.e., two chiral (CW and CCW) edge bands colored in blue and red, spectrally separated by a bulk band due to the existence of synthetic magnetic fields. Approximately ten (set by the lattice dimension) edge resonances within each edge band are represented by another mode number $\sigma$, with a spectral separation of $\approx$~5 GHz near the center of the edge band corresponding to the edge mode round-trip time. \textbf{(B)} Measured (orange) and simulated (grey) integrated dispersion plot for the single ring (left) and the IQH lattice (right). The measurement range covers 25 AGFs (modes $\mu$ from -14 to 11). \textbf{(C)} Representative chip-scale 3D FDTD simulation of field profiles for different topological chiral edge and bulk bands with real device parameters.}
\label{Fig:linear}
\end{figure*}
%
\section{Results and Analysis}
\subsection{Measurement Setup}
The IQH square lattices are implemented in CMOS-compatible ring arrays in an add-drop configuration, with an 800~nm~$\times$~1200~nm SiN waveguide cross-section. The evanescent coupling between adjacent rings, as well as the bus waveguides, uses a directional coupler design, with a coupling length of 12~$\mu$m, and a variety of designed coupling gaps ranging from 300 nm to 600~nm. These gaps are varied to control the coupling strength $J(\mu)$, independent of single-ring geometries, which for these gap values corresponds to a range of 10~GHz to 20~GHz at the pump mode $\mu=0$. The site rings across all designs of IQH lattices are geometrically identical, with an FSR of 773.92~GHz. To realize the multi-modal AGFs in the linear regime via the multi-shot scheme, a continuous-wave laser that is tunable from 183~THz to 202~THz is used, covering 26 site ring modes $\mu$. To realize the ultra-broadband multi-modal AGFs in the nonlinear regime, a narrow-band tunable pulsed laser with a duration of 5 ns is used. The generated IQH topological frequency combs are collected from the drop port and optionally sent to a broadband (125~THz - 250~THz), low-resolution (2.5~GHz) grating-based OSA, or a narrow-band (184~THz - 197~THz), high-resolution (5 MHz) heterodyne-based OSA. To image the intensity profiles of the IQH combs, the out-of-plane scattered light is collected with an infra-red (IR) camera and a 1600~nm long-pass filter above the lattice~\cite{xu2025chip}.
%
\subsection{Multi-shot realization}
We begin with the narrow-band multi-shot realization of the multi-modal AGFs. Fig.\ref{Fig:linear}A shows the drop port spectra of both a single-ring device and an IQH lattice with a gap of 300 nm, measured by pumping the input port with low power (1 mW) and sweeping the frequency over a short range that covers three site ring modes $\mu$. The clear edge-bulk-edge sections in the drop spectra of the lattice showcase the characteristics of the topological band structure of the IQH model~\cite{hafezi2011robust,hafezi2013imaging}. More data and simulations are presented in the SI section~\ref{sm:linear}.

Next, we performed a long-range frequency sweep (183 THz to 202 THz, the full tuning range of our pump) that covers 26 AGFs ($\mu = -14$ to $\mu = 11$), and plot the integrated dispersion $D_{\mathrm{int}}$ from the drop spectra (SI section~\ref{sm:dint_heatmap}), as shown in Fig.~\ref{Fig:linear}B. The measured $D_{\mathrm{int}}$ of both the single ring and the IQH lattice are overlaid with simulations using the dispersive tight-binding model, demonstrating excellent agreement between theory and experiment. In particular, the distinct edge-bulk-edge bands across the entire sweep range for the IQH lattice, in sharp contrast to a conventional AGF picture, exhibit prominent drifts and variations of edge resonances within the edge band as a function of mode number $\mu$, arising from the frequency-dependent AGF captured by our dispersion-corrected theory. We note that a contributing factor to these drifts is the frequency-dependent phase threaded through the IQH plaquette (inset of Fig.~\ref{Fig:intro}F). Specifically, the respective leftward and rightward drifts of the CW and CCW edge modes in the zoomed-in panels of Fig.~\ref{Fig:linear}B can be understood as an analogy to Landau fan where the energy spectrum changes as a function of the magnetic field~\cite{hafezi2011robust,mittal2016measurement}.

To accurately model the spatial profiles of the lattice modes beyond tight-binding approximations, we performed a full chip-scale linear 3D FDTD simulation of the fabricated device, as shown in Fig.~\ref{Fig:linear}C. Selectively exciting different topological bands reveals the distinct field profiles of the two chiral edge bands and the bulk band. Additional 3D FDTD field profile simulations are presented in the SI section~\ref{sm:chipscale}. We note that these time-consuming and computationally challenging full chip-scale simulations are not necessary for our dispersion-corrected AGF framework, which only requires a single unit-cell 3D FDTD simulations of waveguides (SI section~\ref{FigSI: 3D FDTD J}), and is computationally 100 times more efficient.
\begin{figure*}[t]
\centering
\includegraphics[width=0.99\textwidth]{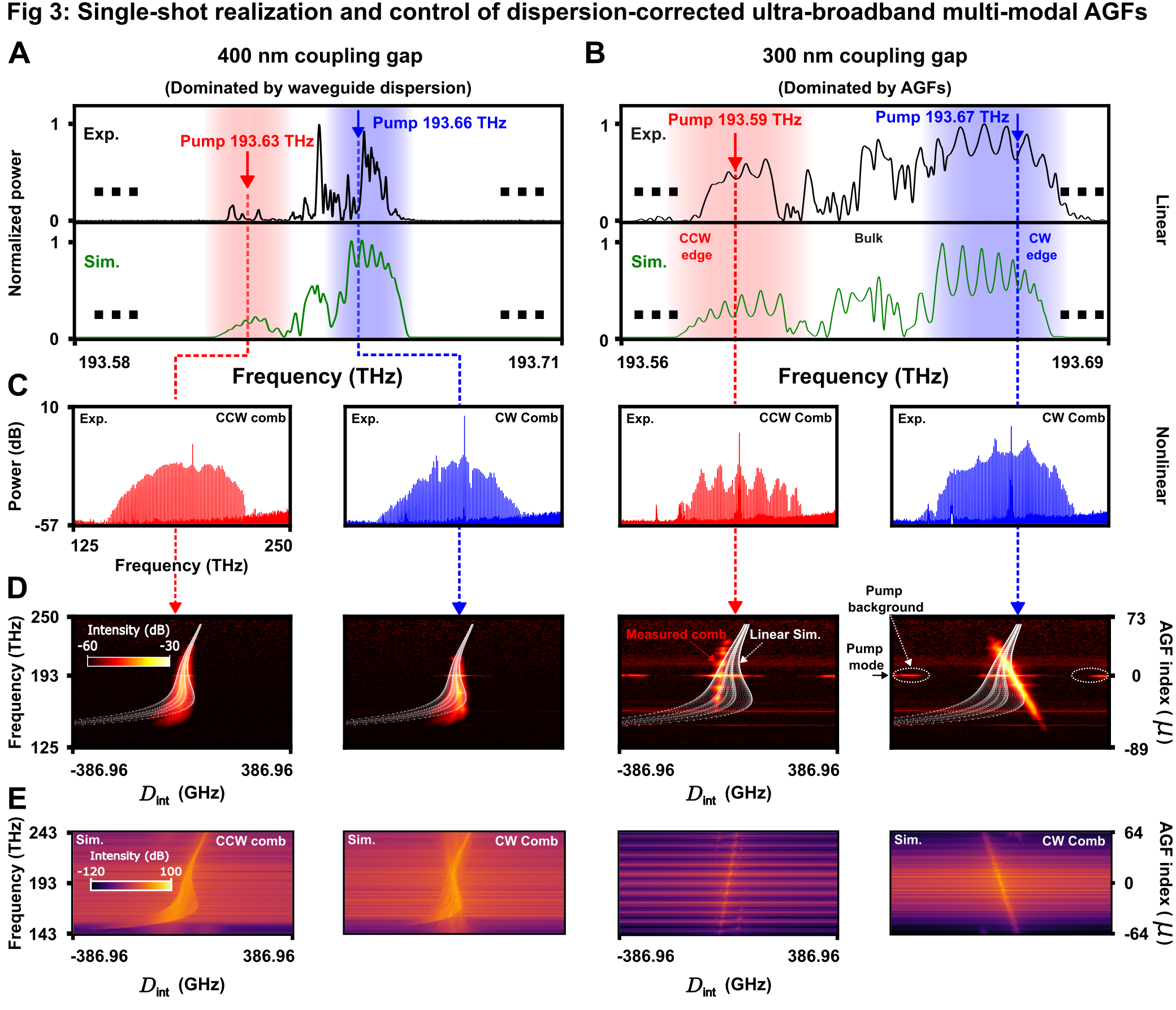}
\caption{Measured (top) and simulated (bottom) drop port linear transmission spectra in the pump band ($\mu=0$) for SiN IQH lattices with a (\textbf{A}) 400~nm and (\textbf{B}) 300~nm coupling gaps, respectively. The chiral CW and CCW edge bands are shaded blue and red, respectively. The 300~nm (400~nm) coupling gap corresponds to a coupling strength $J$ of 20~GHz (10~GHz) at the pump mode. \textbf{(C)} Comb spectra and \textbf{(D)} the corresponding integrated dispersion plots (red heatmap) for CW and CCW edge excitation frequencies marked by dashed blue and red lines in (A,B), respectively. The simulated linear regime AGF landscapes (white dots) are overlaid with nonlinear experiments in (D) for comparison. \textbf{(E)} Simulated nonlinear-regime integrated dispersion plots corresponding to (D). For comb generation, a pump power of $\approx$~233~mW ($\approx$~138~mW) is used for the 300~nm (400~nm) coupling gap device.}
\label{Fig:generate_comb}
\end{figure*}
%
\subsection{Single-shot realization}
To realize ultra-broadband multi-modal AGFs with a single shot, we pumped the IQH lattice around the center of edge bands with a high-power tunable pulsed laser with a duration of 5 ns and a repetition rate of 250 kHz (operating in the quasi-continuous-wave pumping regime~\cite{xu2025chip}). At high powers, cascaded four-wave mixing forms a comb with an OPO threshold of $\approx$~200~mW (SI section~\ref{sm:power}). The comb bandwidth grows with power, spanning nearly an octave. Figures~\ref{Fig:generate_comb}A,B,C show the frequency comb generation in IQH lattices with coupling gaps of 400~nm and 300~nm by pumping two frequencies in different chiral edge bands of the mode $\mu=0$. 

The integrated dispersion $D_{\mathrm{int}}$ of the corresponding combs is shown in Fig.~\ref{Fig:generate_comb}D, and overlaid with the linear integrated dispersion predicted by the dispersive tight-binding Hamiltonians. We observe that, for the lattice with a gap of 400 nm, the generated modulation instability frequency comb largely follows the linear dispersion as it fills each resonance with nonlinearly generated photons. 
In contrast, for the 300~nm-gap lattice, the generated frequency combs do not follow the integrated dispersion and instead form straight lines that indicate an equidistant grid of single-ring frequency markers (in the $\mu$-space), a signature akin to a mode-locked state.
Moreover, the resulting ``straight-line'' comb spectrum exhibits a slope in the frequency-$\mu$ space due to a repetition rate difference from the linear $D_1$ at the pump frequency. Additionally, we note that the slope's sign and magnitude depend strongly on the pump mode as well as the edge band, yielding a deviation of -1.66~GHz to 0.35~GHz from the linear $D_1=773.92$~GHz.
We note that, while this slope should correspond to the comb's mode-locking in the fast time scale~\cite{xu2025chip} (\textit{i.e.}, relative to the individual site ring), it appears directly correlated to the dispersion of the edge mode (Fig.\ref{Fig:intro}), highlighting in this higher tunneling regime the interlinking of slow and fast time-scales.
Importantly, the edge-mode dispersion is highly dispersive near the band gap, providing a large variation of the repetition rate against the linear FSR value.
This behavior resembles the slow-light regime in photonic crystal waveguides, where the enhancement in the density of states results in a reduction of the threshold power of comb generation. Additional results on the pump power dependence and the multi-harmonic generation are shown in the SI sections~\ref{sm:power} and~\ref{sm:image}. Information on the comb generation on a single-site ring is presented in the SI section~\ref{sm:singlering}.
These observations are accurately captured by the dispersion-corrected nonlinear simulations in Fig.~\ref{Fig:generate_comb}E from our dispersive tight-binding model. We note that the only difference between lattices in Figs.~\ref{Fig:generate_comb}A,B is the ring-to-ring coupling gap; therefore, our results indicate lattice-scale dispersion engineering beyond the waveguide dispersion of the waveguide. 
\begin{figure*}[t]
\centering
\includegraphics[width=0.99\textwidth]{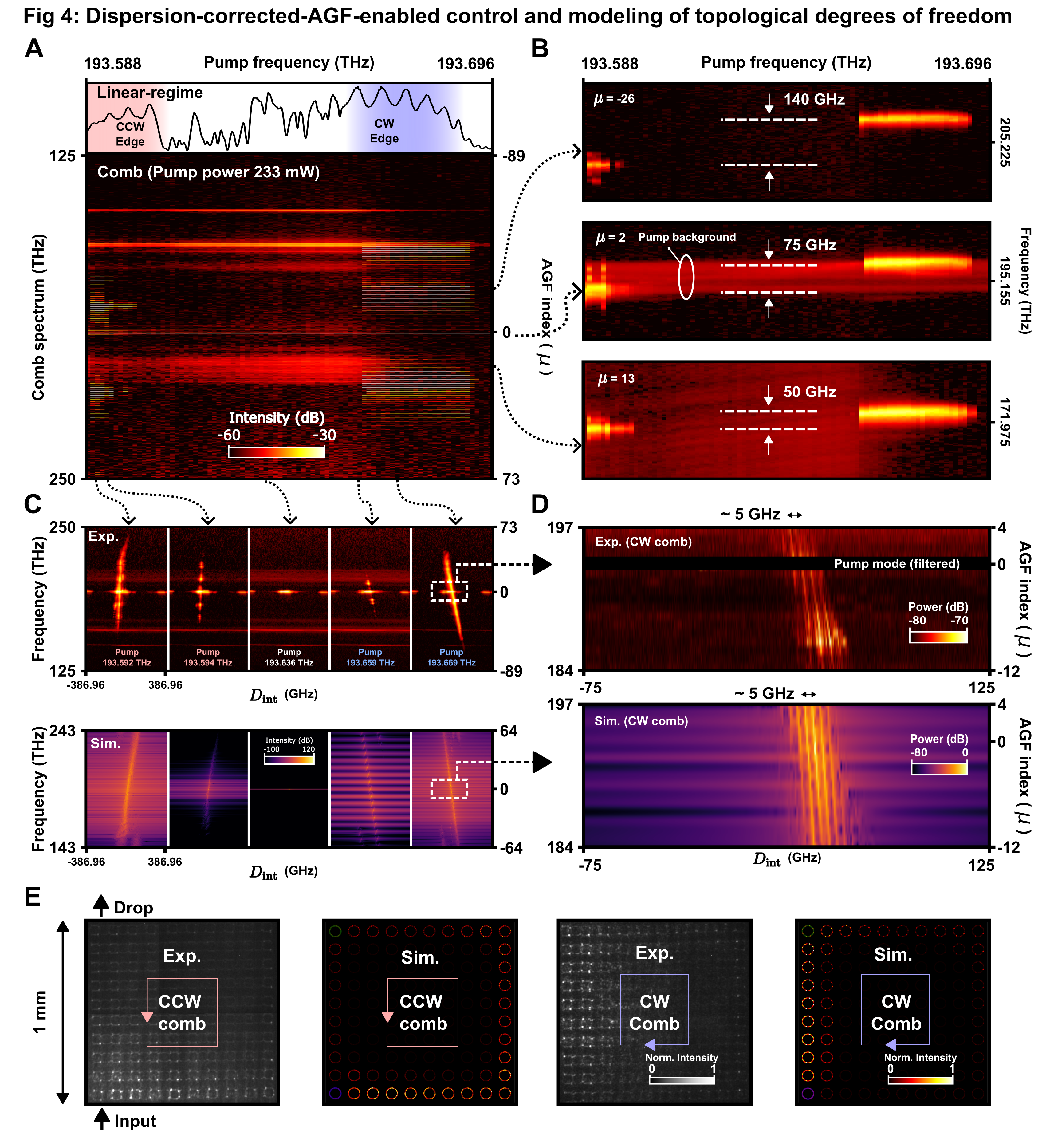}
\caption{\textbf{(A)} Pump frequency–dependent comb spectra from the drop port of a 10~$\times$~10 SiN IQH lattice, pumped with a fixed power of 233 mW. The pump is swept through the topological band $\mu=0$, with chiral CW and CCW edge bands shaded blue and red in the linear-regime spectra, respectively (top panel). No combs are generated in the bulk band due to the high OPO threshold. \textbf{(B)} Zoomed in comb teeth spectra, corresponding to AGFs $\mu=-26,2,13$, showing an $\mu$-dependent edge-edge band gaps marked with white dashed lines. \textbf{(C)} Measured (top row) and simulated (bottom row) integrated dispersion plots for five different pump frequencies, exhibiting opposite comb dispersion slopes for opposite edge combs. \textbf{(D)} Heterodyne-based nesting-resolved measured (top row) and simulated (bottom row) integrated dispersion plots of a representative CW-edge IQH comb shown in (C), covering modes $\mu$ from -12 to 4. The 5~GHz nested mode spacing is marked. \textbf{(E)} Measured and simulated comb intensity profiles for a representative CW and CCW edge comb, respectively. The pump is filtered in all cases. }
\label{Fig:nonlinear_topology}
\end{figure*}
%
\subsection{AGF-enabled control over spatial-spectral properties and nested structure}
To demonstrate control mechanisms enabled by the AGFs over the spatial and spectral degrees of freedom of the IQH lattice in the nonlinear regime, we used a fixed pump power above the OPO threshold at approximately $233~\mathrm{mW}$, and swept the pump frequency across the entire topological band at $\mu=0$. The results are summarized in Fig.~\ref{Fig:nonlinear_topology}, which shows AGF-enabled comb spectra and spatial profile characteristics. We performed a focused analysis of three representative comb teeth (Fig.~\ref{Fig:nonlinear_topology}B). For all selected modes, the comb teeth generated by pumping different chiral edge bands of mode $\mu=0$  are spectrally separated, with spacings ranging from $140~\mathrm{GHz}$ for $\mu=-26$, to $75~\mathrm{GHz}$ for $\mu=2$, and $50~\mathrm{GHz}$ for $\mu=13$. These observations have several important implications. First, the varying separation directly reflects the strongly dispersive characteristic of the AGFs in agreement with our dispersion-corrected AGF theory. Moreover, the persistence of this separation across all the selected AGFs demonstrates that the topological band structure of the IQH lattice--where a bulk band separates two edge bands--remains robust across multiple modes. Furthermore, the fact that pumping one chiral edge band does not generate comb light in the opposite edge band or the bulk band, with an extinction ratio of up to $30~\mathrm{dB}$, indicates strong suppression of CW-CCW backscattering as well as edge-bulk mode mixing. This serves as direct evidence for the preservation of topological protection for light even in the presence of strong Kerr nonlinearity.

Next, we performed integrated dispersion analysis on frequency combs generated when pumping either chiral edge band. On one hand, the bandwidth of combs in either edge band approaches an octave. Moreover, as the pump frequency is swept from CW to the CCW edge band, the integrated dispersion profiles evolve, including a clear sign-flip of the comb dispersion's slope. This observation, which is in excellent agreement with dispersion-corrected AGF comb simulations (Fig.~\ref{Fig:nonlinear_topology}C), demonstrates a flexible control mechanism over comb spectrum beyond the single-ring waveguide dispersion (SI Section~\ref{sm:fullsweep}).
\begin{figure*}[t]
\centering
\includegraphics[width=0.8\textwidth]{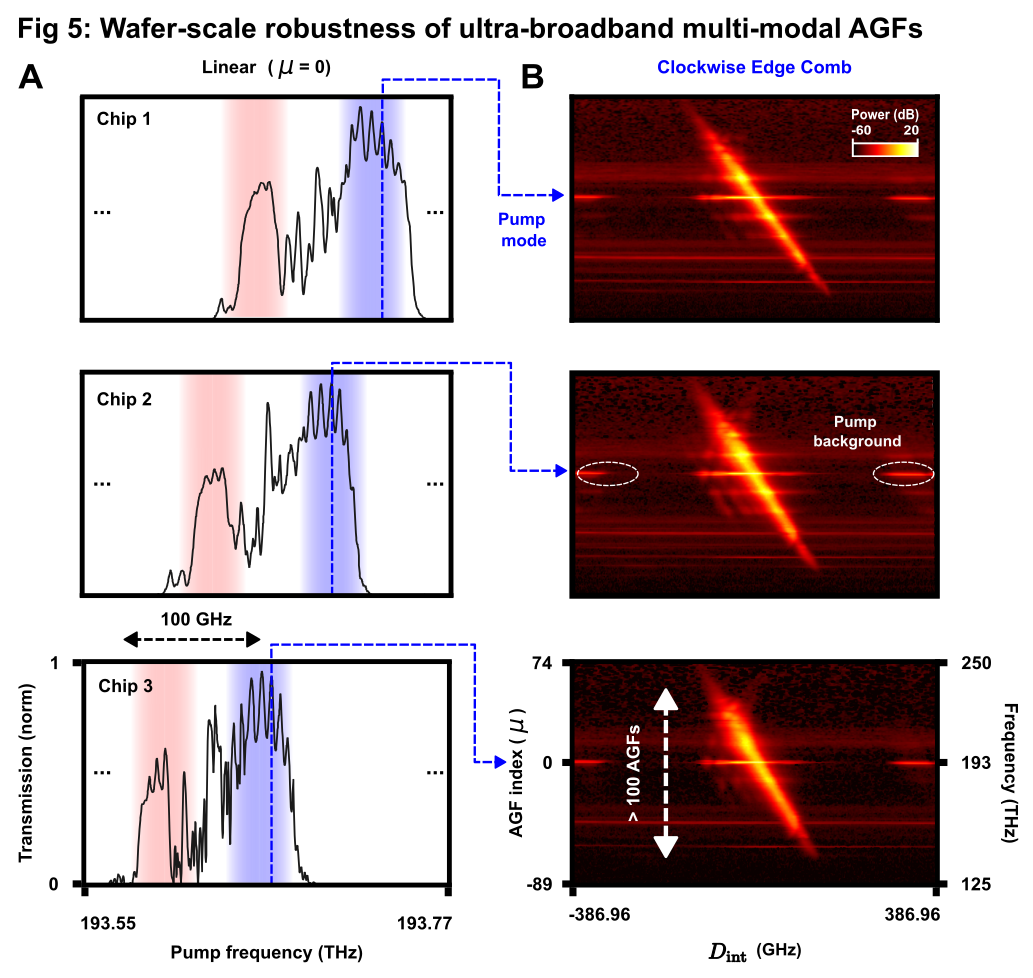}
\caption{\textbf(A) Drop port linear transmission spectra in the pump band ($\mu=0$) and \textbf(B) the corresponding integrated dispersion plots of CW IQH combs from three identically designed IQH devices with a 300~nm coupling gap on separate chips, with chiral CW and CCW edge bands shaded blue and red, respectively. The 300~nm coupling gap corresponds to coupling strength J of $\approx$~20~GHz at the pump band. 100~GHz chip-to-chip spectral shifts in (A), induced by fabrication variation, are marked. For each device, the pump frequency for comb generation is marked by vertical dashed lines in the drop spectra.}
\label{Fig:wafer_scale}
\end{figure*}

To resolve the nested structure of multi-modal AGFs in the nonlinear regime, we used a high-resolution heterodyne-based OSA to inspect the IQH comb presented in Fig.\ref{Fig:nonlinear_topology}C. Fig.~\ref{Fig:nonlinear_topology}D plots the measured and simulated integrated dispersion of the CW edge comb, which clearly shows the well-defined nested structures of the comb teeth with a characteristic spacing of 5~GHz, in agreement with linear-regime measured edge mode spacing as well as the nonlinear simulation. Furthermore, comparison with the linear response confirms that pumping the CW chiral edge does not generate light in the CCW chiral edge band, indicating the absence of backscattering between the chiral edge states in the nonlinear regime (SI section~\ref{sm:apex}). Moreover, numerical simulations demonstrate that nested topological solitons~\cite{mittal2021topological} can exist in the IQH lattice with realistic parameters (SI section~\ref{sm:LLE_soliton}).  Importantly, such nested edge combs, where more than 100~AGFs (10000 photonic modes) approach an octave in bandwidth, each with a nested set of up to $\approx$~10 comb teeth (set by the lattice dimension), can potentially comprise up to $\approx$~1000 Kerr comb lines. However, we note that confirmation of the exact number of total comb lines requires a broader and more sensitive heterodyne-based OSA, which may be an intriguing future study.

To assess the AGF-enabled control over the spatial degrees of freedom of the IQH combs, we imaged the spatial intensity profiles of the combs generated in both chiral edge bands. We collected the light scattered out of the plane above the lattice with an IR camera and a 1600 nm long-pass filter to exclusively isolate the comb light from the pump. Fig.~\ref{Fig:nonlinear_topology}E shows the IR imaging results and corresponding simulations of the comb intensity profiles. Controlled by the edge-band-selective excitation frequency, the edge-confined comb light circulates in the opposite direction in the IQH lattice, remaining robust around sharp 90° turns with negligible bulk leakage. Moreover, in sharp contrast to conventional single-rings~\cite{yang2017counter} as well as previous anomalous quantum Hall combs~\cite{flower2024observation,xu2025chip}, the IQH lattice enables the control of the edge combs' chirality simply by changing the pump frequency (by $\approx$~50~GHz) while pumping the same input port.

%
\subsection{Wafer-scale robustness}
To assess the robustness of AGFs against wafer-scale fabrication variations, we measured multiple devices with identical designs fabricated on separate chips, including both the 300~nm and 400~nm coupling gap configurations. The results are summarized in SI section~\ref{sm:wafer}. For the 300~nm cases, the linear drop port spectra (Fig.~\ref{Fig:wafer_scale}A) show 100~GHz chip-to-chip spectral shifts and profile variations, which are due to wafer-thickness and disorder variations. Remarkably, despite these shifts, using comparable pump power, all devices consistently exhibit single-shot realization of more than 100~AGFs approaching an octave in bandwidth (Fig.~\ref{Fig:wafer_scale}B). Moreover, the dispersion profiles of the IQH combs for the 300~nm coupling gap configuration remain in excellent agreement with the dispersion-corrected nonlinear theory predictions in Fig.~\ref{Fig:generate_comb}D for each case.
%
\section{Summary and outlook}
Our single-shot-realizable, dispersion-correction framework for ultra-broadband, multi-modal AGFs --valid from linear response to nonlinear dynamics--addresses a central limitation of conventional single-mode tight-binding descriptions: outside a narrow-band regime, the gauge landscape ceases to be well defined. By restoring a faithful gauge description when dispersion and mode dependence are unavoidable, our approach extends AGF physics deep into the ultra-broadband, multi-modal domain. Beyond the IQH comb demonstrated here, the rapidly expanding field of AGF-based frequency combs in integrated micro-rings can benefit directly from dispersion-correction, including anomalous~\cite{mittal2021topological}, Floquet~\cite{hashemi2024floquet}, non-Hermitian~\cite{hashemi2025reconfigurable}, hyperbolic~\cite{huang2024hyperbolic}, and hierarchically nested~\cite{mehrabad2025quantum} topological lattices. Experimentally, this framework enables quantitative modeling of anomalous quantum Hall nested frequency combs~\cite{flower2024observation}, multi-timescale synchronizers~\cite{xu2025chip}, and high-harmonic processes~\cite{mehrabad2025multi}, where existing descriptions remain incomplete. Similarly, slow-light-enhanced comb generation schemes in coupled ring arrays can benefit from our scheme~\cite{vasco2019slow}. In the linear regime, higher-order phases~\cite{mittal2019photonic}, mixed-topology lattices~\cite{mehrabad2025quantum}, and spin~\cite{yang2021optically} and valley Hall~\cite{liu2025winding} phases can similarly be extended far beyond narrow-band operation.

More broadly, while chromatic dispersion has traditionally limited AGFs--particularly in synthetic frequency lattices where dissipation or frequency walk-off disrupts lattice connectivity~\cite{DuttNatCommun2022,HuNatCommun2022}--our work re-frames dispersion as a controllable design degree of freedom. Building on established concepts in unconventional coupling engineering~\cite{ChinJ.LightwaveTechnol.JLT1998,HosseiniOpt.ExpressOE2010,MoilleOpt.Lett.2019a,Kim2025} and photonic inverse design~\cite{MoleskyNaturePhoton2018,AhnACSPhotonics2022}, frequency-dependent dissipation and coupling can be tailored to open multiple, well-separated edge-band windows across the spectrum, enabling multiplexed AGFs with increased functional capacity. Within this unified framework, on-site potentials, coupling strengths, and hopping phases are treated as frequency-dependent quantities extracted from broadband 3D FDTD, enabling gauge-field engineering that goes fundamentally beyond waveguide dispersion while remaining robust against wafer-scale fabrication variations. Combined with single-shot nonlinear probing and wafer-scale reproducibility, this approach provides a scalable route to dispersion-resilient, volume-manufacturable AGF-based photonic circuits operating far beyond the single-mode limit, including reconfigurable lattice-scale dispersion engineering~\cite{dai2024programmable} for integrated Ising machines~\cite{al2025programmable} as well as dynamical lattice gauge theories~\cite{del2023dynamical}, in particular for strongly coupled ring lattices in which light intensity in link rings is strong~\cite{hashemi2024floquet}.

Finally, dispersion-corrected ultra-broadband AGF photonics opens a pathway toward programmable optical simulation~\cite{suh2024photonic,liao2025hetero}, memory~\cite{ahmadnejad2025nontrapping}, computing, and artificial intelligence. Narrow-band, linear-regime AGFs have already been explored for on-chip photonic processing~\cite{cataldo2025topological}, classification~\cite{du2024ultracompact}, interconnects~\cite{wang2024ultralow}, beamformers and optical links~\cite{wang2024chip}, and highly reconfigurable routing~\cite{dai2024programmable,ma2024anisotropic} and simulators~\cite{dong2025sl}. By lifting bandwidth and single-mode constraints while maintaining robustness against disorder, our framework substantially expands this toolbox, enabling new classes of dispersion-corrected dynamics, simulations, and device-level functionalities that unify nonlinear and quantum photonics on chip.
%

\section{Acknowledgments}
The authors acknowledge fruitful discussions with Sunil Mittal, Yanne Chembo, Avik Dutt, Chao Li, Zhiquan Yuan, and Dirk Englund. M.J.M., L.X., S.S., and M.H. acknowledge Flexcompute for access to simulation resources based on their Tidy3D software. Certain commercial equipment, instruments, or materials (or suppliers, software, etc.) are identified in this paper to foster understanding. Such identification does not imply recommendation or endorsement by the National Institute of Standards and Technology, nor does it imply that the materials or equipment identified are necessarily the best available for the purpose. This work was supported by DARPA DSO HR0011-26-3-E053 and ARO W911NF261A018.
%
\section{Author contributions}
All authors have contributed to this work.
%
\section{Competing interests}
M.J.M., L.X., A.P., S.S., and M.H. have filed a patent covering the realization of dispersion-corrected ultra-broadband multi-modal AGFs. The authors declare no other competing interests.
%
\section{Data availability}
All of the data that support the findings of this study are reported in the main text and Supplementary Information. All data are available in the manuscript, the supplementary material or deposited at Zenodo~\cite{dataset}.
%
\newpage

\section*{Supplementary Information for\\ 
Single-Shot Realization of 10000-Mode Octave-Spanning Artificial Gauge Fields}


\section*{Supplementary Table of Contents}
\vspace{-0.75em}
\hrule
\vspace{0.9em}

\newcommand{\SupTOCEntry}[3]{%
  #1.\quad\hyperref[#2]{#3}%
  \dotfill\pageref{#2}\par
}

\begingroup
\setlength{\parindent}{0pt}
\setlength{\parskip}{0.35em}


\SupTOCEntry{S1}{sm:setup}
  {Device fabrication and experimental setup}
\SupTOCEntry{S2}{sm:AQHIQH}
  {Comparison between AQH and IQH lattices}
\SupTOCEntry{S3}{sm:dtb}
  {Dispersive tight-binding Hamiltonians}
\SupTOCEntry{S4}{sm:2ddispersion}
  {Two-dimensional integrated dispersion grid and lattice-scale dispersion engineering}
\SupTOCEntry{S5}{sm:lle}
  {Lugiato-Lefever simulations of the IQH combs}
\SupTOCEntry{S6}{sm:linear}
  {Linear drop port measurement and simulation of IQH lattices}
\SupTOCEntry{S7}{sm:dint_heatmap}
  {Constructing the integrated dispersion plot}
\SupTOCEntry{S8}{sm:chipscale}
  {Chip-scale 3D FDTD simulations of the drop spectrum and field profile of IQH lattices.}
\SupTOCEntry{S9}{sm:power}
  {Pump power dependence of comb spectra and integrated dispersion}
\SupTOCEntry{S10}{sm:image}
  {Spatial imaging of generated harmonics}
\SupTOCEntry{S11}{sm:singlering}
  {Comb generation and integrated dispersion for a single ring resonator}
\SupTOCEntry{S12}{sm:fullsweep}
  {Full sweep simulation of IQH comb dispersion}
  \SupTOCEntry{S13}{sm:apex}
  {High resolution optical spectrum measurements}
\SupTOCEntry{S14}{sm:LLE_soliton}
  {Topological IQH solitons over multi-modal AGFs}

\SupTOCEntry{S15}{sm:wafer}
  {Wafer-scale reproducibility}

\SupTOCEntry{Supplementary Movie M1}{sm:LLE_soliton}
{Video of an IQH topological soliton}
 
\SupTOCEntry{Supplementary Movie M2}{sm:LLE_soliton}
{Video of an IQH topological soliton with a defect}

\begin{supplement}
\newpage

%
\section{Device fabrication and experimental setup}\label{sm:setup}
A schematic of the experimental setup and a high-resolution optical image of a representative fabricated photonic topological IQH lattice are shown in Fig.~\ref{FigSI: setup}. The devices are fabricated at a commercial foundry, following a similar process as described in Refs.~\cite{flower2024observation,xu2020photonic,mehrabad2025multi}. The micro-ring resonators are made of Si$_3$N$_4$ waveguides with an 800~nm by 1200~nm cross-section and capped with a SiO$_2$ cladding.

For the linear regime measurements, a low-power (1~mW) continuous-wave tunable laser is coupled into the input port using edge couplers, and the wavelength is continuously swept across a broad range of single-ring resonances. Simultaneously, the output power at the drop port across the lattice is measured using a power meter. To determine the group delay of the transmission at the drop port, an optical vector network analyzer is used. Polarization control in both setups is managed using a standard 3-paddle polarization controller placed after the laser output.

For nonlinear measurements, a high-power (up to 700~mW average power, capable of reaching above the OPO threshold) pulsed tunable laser is directed into a free-space optical arrangement that includes a variable attenuator and a polarization controller consisting of a quarter-wave plate, a half-wave plate, and another quarter-wave plate. The output is subsequently coupled into a short tapered fiber and edge-coupled into the SiN chip via the input port. The coupling efficiency for each coupler is estimated to be between -2~dB and -3~dB. The combined pump and comb output is collected from the drop and through ports using another tapered fiber and is optionally attenuated or filtered before detection. The laser used for comb generation is a 5~ns pulsed laser with a repetition rate of 250~kHz, tunable from 1546~nm (193.91~THz) to 1552~nm (193.17 THz), covering a range comparable to one FSR of the single-ring.

For spatial imaging, out-of-plane scattering from the SiN chip is captured using a 10x objective lens with a numerical aperture of 0.28. The image is directed through a 50:50 beamsplitter, allowing it to be captured by both a visible wavelength camera and an infrared (IR)-sensitive camera. A 1580~nm (189.74~THz) long-pass filter is positioned in front of the IR camera to exclude the pump laser.

\begin{figure*}[h]
\centering
\includegraphics[width=1\textwidth]{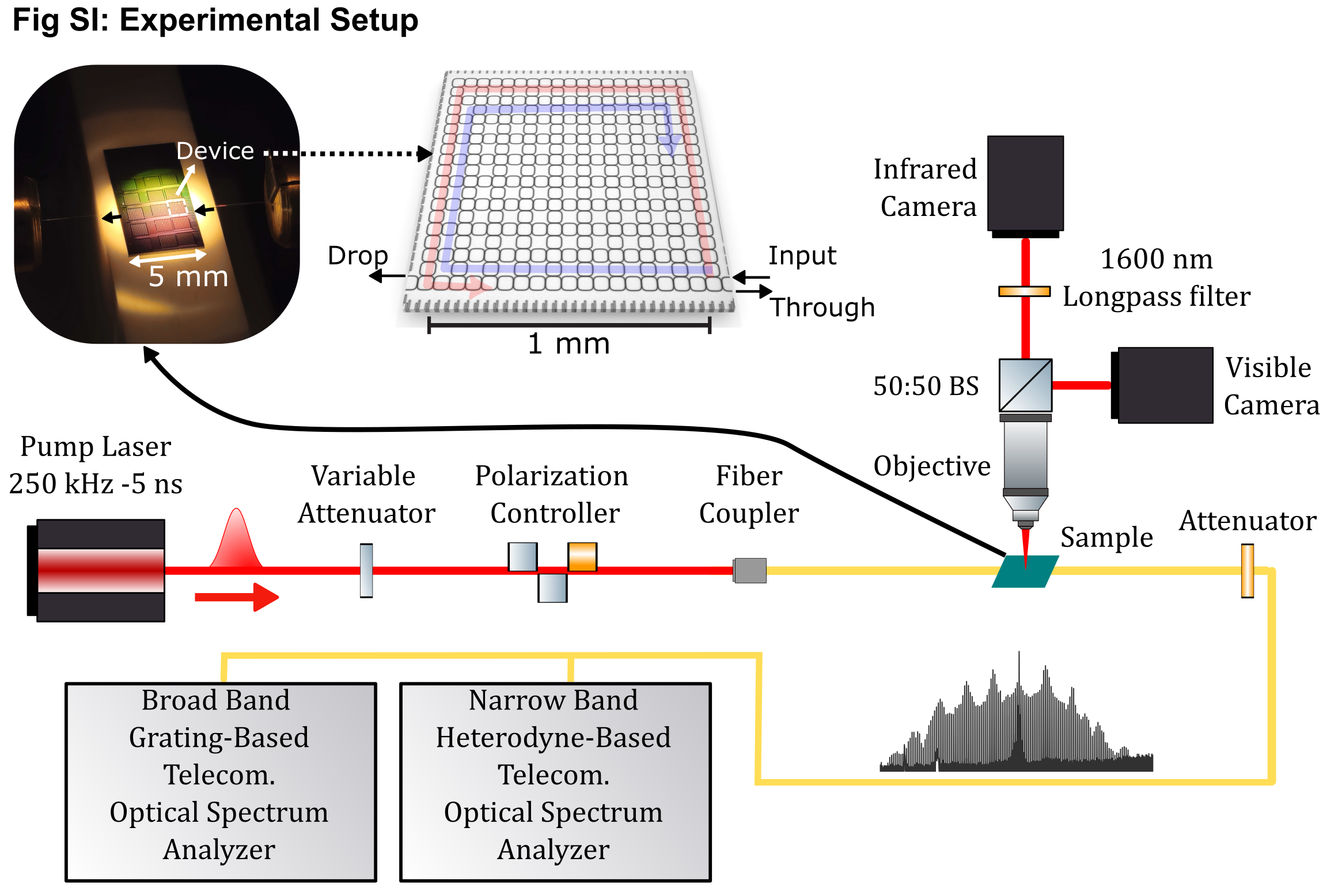}
\caption{A tunable telecom (near~1550~nm) pulsed laser with a repetition rate of 250~kHz and pulse duration of 5~ns is attenuated and polarization-controlled before coupling into the lattice. A photo of the chip captured by a smart phone, as well as a high-resolution optical microscope image of the lattice, with its input/output/drop ports marked, and schematized chiral edge states (blue and red), are shown. An objective lens above the sample is used to collect light scattered out of the lattice's plane. The light is split by a 50:50 beamsplitter and imaged using a visible camera and an infrared camera. The generated topological frequency comb collected from the through/drop port is optionally sent to optical spectrum analyzers with different bandwidths and resolutions. }
\label{FigSI: setup}
\end{figure*}
\newpage
%
\section{Comparison between AQH and IQH lattices}\label{sm:AQHIQH}
Recent experimental studies on topological frequency combs were conducted on anomalous quantum Hall (AQH) lattices\cite{flower2024observation,xu2025chip,mehrabad2025multi}, which are different from the IQH lattices studied in this work. In this section, we highlight the key differences between these two different topological photonic models, focusing on a narrow-band near mode $\mu=0$, as summarized in Fig.~\ref{FigSI: aqh_iqh}. 

Fig.~\ref{FigSI: aqh_iqh}A shows the IQH lattice. The plaquette of an IQH consists of 4 site rings connected by 4 link rings that are anti-resonant relative to the site rings. The anti-resonant condition is realized via a 440~nm longer optical path length for the link rings. Moreover, the link rings are spatially configured 
(here by shifting the link rings by a small amount in the y direction) such that a synthetic magnetic field simulating the Landau gauge is realized. For a lattice with size $N$, i.e., $N$ site rings on each side of the square lattice, the location of a site ring is labeled by a Cartesian coordinate (x,y), and the corresponding single-mode Hamiltonian of an IQH lattice near $\mu=0$ reads:
\begin{equation}
\begin{aligned}
H_{\mathrm{IQH}}
= \quad \sum_{\substack{x,y}}\omega_0 \hat{a}^{\dagger}_{x,y}\,\hat{a}_{x,y}-\, J \sum_{\substack{x,y}}
\Big( \hat{a}^{\dagger}_{x+1,y}\,\hat{a}_{x,y}\,
e^{-i\,y\,\phi} \;+\;
\hat{a}^{\dagger}_{x,y+1}\,\hat{a}_{x,y} \;+\; \mathrm{h.c.} \Big)
\end{aligned}
\label{eqn:single_mode_tb_H_iqh}
\end{equation}
\noindent where $J$ is the effective evanescent coupling strength between two site-rings, and $\phi$ is the phase acquired by a photon when it completes a loop in the plaquette, determined by the choice of the synthetic magnetic field. In our case, the Landau gauge is used for the synthetic magnetic field; therefore, a photon acquires a phase only when it hops along the x direction~\cite{hafezi2011robust,hafezi2013imaging}. 

Fig.~\ref{FigSI: aqh_iqh}B shows the AQH device. The plaquette of an AQH lattice consists of 4 site rings and 1 link ring, where, in contrast to the IQH case, no spatial shift for the link rings is introduced. Using a single number (m or n) to represent the location of a site ring, the Hamiltonian of an AQH lattice reads~\cite{leykam2016anomalous}:
\begin{equation}
H_{AQH}
= \sum_{\substack{m}}\omega_0 \hat{a}^{\dagger}_{m}\,\hat{a}_{m}-J \sum_{\langle m,n\rangle} 
a_{m}^\dagger a_{n} e^{-i\phi_{m,n}}
- J \sum_{\langle\!\langle m,n\rangle\!\rangle}
a_{m}^\dagger a_{n}
+ \text{h.c.}
\end{equation}
\noindent where $\langle\!\langle m,n\rangle\!\rangle$ represents two-site rings labeled by m and n that are next-nearest neighbors, while $\langle m,n\rangle$ represents nearest neighbors.
Fig.~\ref{FigSI: aqh_iqh}C,D compare the spectral signatures, the photon flow, and intensity profiles for the IQH and AQH lattices. Here, the photon flow for a given mode $\mu$ and site ring coordinate (x,y) is calculated by:

\begin{equation}
\begin{aligned}\rm{Flow}(x,y;\mu)&=(<H_{(x,y)}^{(x+1,y)}(\mu)\;\hat{a}^{\dagger}_{x+1,y}(\mu)\,\hat{a}_{x,y}(\mu)>-<H_{(x-1,y)}^{(x,y)}(\mu)\;\hat{a}^{\dagger}_{x,y}(\mu)\,\hat{a}_{x-1,y}(\mu)>)\hat{x}\\&\quad+(<H_{(x,y)}^{(x,y+1)}(\mu)\;\hat{a}^{\dagger}_{x,y+1}(\mu)\,\hat{a}_{x,y}(\mu)>-<H_{(x,y-1)}^{(x,y)}(\mu)\;\hat{a}^{\dagger}_{x,y}(\mu)\,\hat{a}_{x,y-1}(\mu)>)\hat{y}
\end{aligned}
\end{equation}

\noindent Here $H_{(x,y)}^{(x+1,y)}(\mu)$ is the element of the Hamiltonian $\mathcal{H}(\mu)$ that connects two site rings with coordinates $(x,y)$ and $(x+1,y)$. For the IQH lattice, two edge bands, shaded blue and red, are spectrally separated, whereas for an AQH lattice, a single edge band is sandwiched between two bulk bands. Notably, the AQH model arises without the need for an external gauge field (the total flux is zero), in contrast to the IQH case, which requires a non-zero external synthetic magnetic field to induce Landau level quantization.
\begin{figure*}[h]
\centering
\includegraphics[width=1\textwidth]{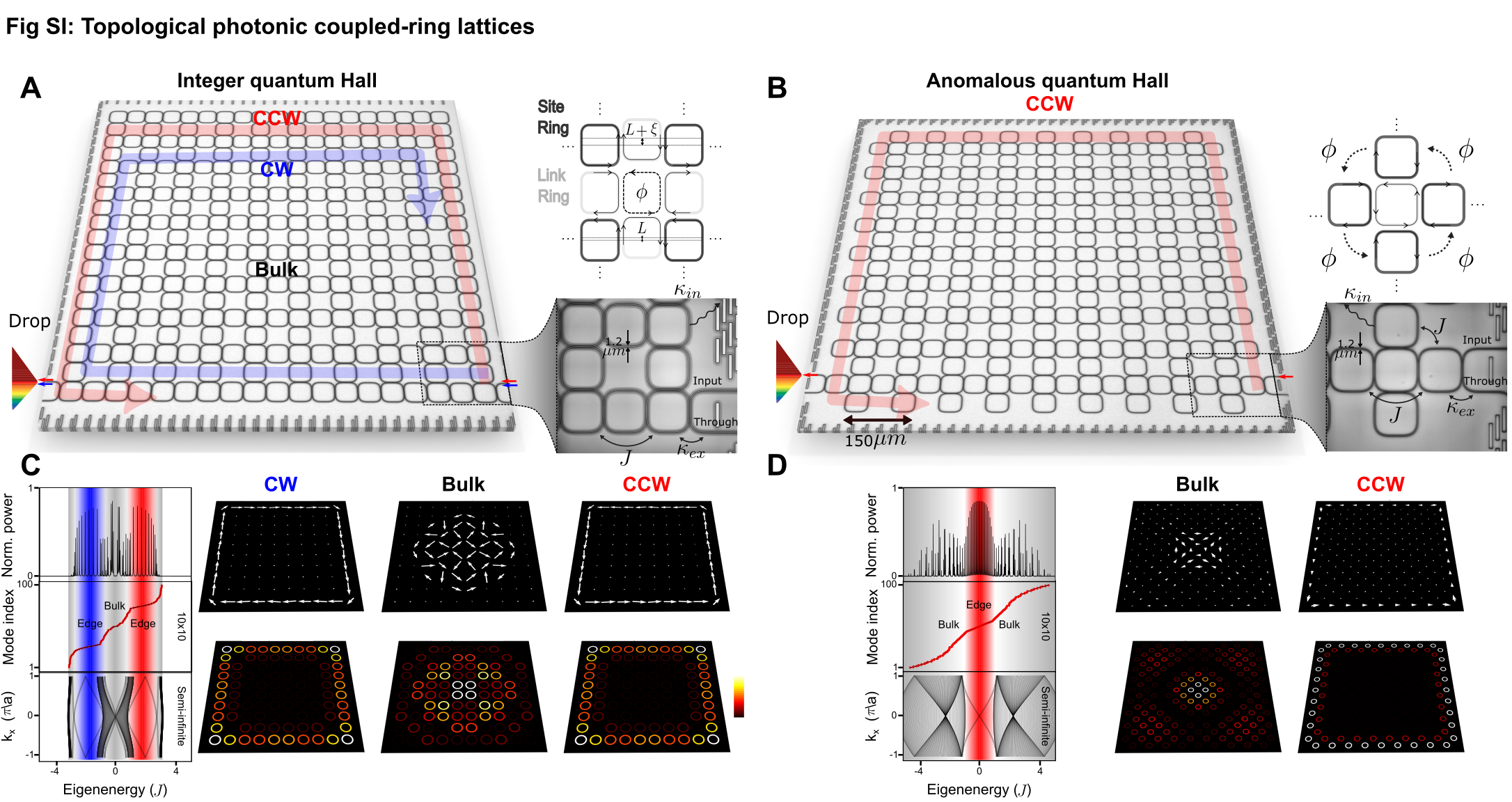}
\caption{\textbf{(A)} High-resolution image of a representative topological IQH lattice. A plaquette, as well as a bus waveguide, is shown on the right. In the IQH plaquette, link rings are spatially shifted to introduce a synthetic magnetic field. By pumping the same port on the lattice with different pump frequencies, chiral edge states can be selectively excited, as marked via the clockwise (CW) and counter-clockwise (CCW) arrows. \textbf{(B)} High-resolution image of a representative topological AQH lattice. The plaquette of an AQH lattice is shown on the right, where the link ring is not shifted. From one input port, only one helical edge state can be excited. \textbf{(C)} Left: from top to bottom, the linear drop spectra, the eigenvalues (10~$\times$~10 lattice), and the band structure (for a semi-infinite) of an IQH lattice as a function of eigenenergy are shown for a narrow-bandwidth near mode $\mu=0$. Right: representative photon flow and intensity profile plots for an IQH lattice. The two chiral edge bands are shaded blue and red, respectively. \textbf{(D)} Same plots as \textbf{(C)}, but for an AQH lattice.}
\label{FigSI: aqh_iqh}
\end{figure*}
\newpage
%
\section{Dispersive tight binding Hamiltonians}\label{sm:dtb}
In this section, we present more details for the dispersion-corrected, i.e., dispersive, tight-binding framework introduced in this work. To start with, we note that the IQH Hamiltonian~\ref{eqn:single_mode_tb_H_iqh} is valid only for narrow-band, single-mode regime, whose validity beyond this regime is constrained by four conditions. First, the link rings need to remain anti-resonant with respect to the site rings. Second, the frequency range needs to be narrow-band, such that the frequency-dependence of $J,\phi$ is negligible. Third, only a single-mode approximation is used, such that the waveguide dispersion of micro-ring resonators can be ignored. Finally, it is important that the coupling strength $J\ll \rm{FSR}$~\cite{hashemi2024floquet}, such that each resonator can be treated as an individual lattice site. However, in this study, as well as recent works on nonlinear topological photonics~\cite{flower2024observation,xu2025chip,mehrabad2025multi}, where frequency combs approach broadband widths with many modes, the frequency-dependence of $J,\phi$, and the integrated dispersion of site rings can no longer be neglected. 

One method to address these frequency-dependent complications is to use broadband 3D FDTD simulations for the lattice. However, a full 3D FDTD simulation of the entire lattice (typically 1 mm~$\times$~1~mm) across an octave (100~THz) is computationally very challenging. Nevertheless, one can employ a hybrid method by treating the two scales (single-ring and edge channel round-trip times) separately. With this spirit, one can simply perform 3D FDTD simulations of a waveguide and a directional coupler, while still treating each ring as a point on the lattice level, which leads to a dispersive tight-binding Hamiltonian that reads: 
\begin{equation}
\begin{aligned}
H_{\mathrm{IQH}}(\mu)
&= \sum_{x,y\le N} \,\{D_{\mathrm{int}}(\mu)-J(\mu)\rm{cot}[\beta(\mu)\eta/2]\}\;
\hat{\it{a}}^{\dagger}_{x,y}(\mu)\,\hat{\it{a}}_{x,y}(\mu) \\
&\quad -\, J(\mu)/\rm{sin}[\beta(\mu)\eta/2] \sum_{\substack{x,y\le N}}
\Big( \hat{\it{a}}^{\dagger}_{x+1,y}(\mu)\,\hat{\it{a}}_{x,y}(\mu)\,
e^{-i\,y\,\phi(\mu)} \;+\;
\hat{\it{a}}^{\dagger}_{x,y+1}(\mu)\,\hat{\it{a}}_{x,y}(\mu)\;+\; \mathrm{h.c.} \Big).
\end{aligned}
\label{eqn:dispersive_tight_binding}
\end{equation}
This corresponds to the $H_\text{IQH}(\mu)$ defined in the main text with $\zeta \equiv \eta/2x$. The parameters are introduced in the following paragraghs.

To provide insights for equation~\ref{eqn:dispersive_tight_binding}, we emphasize that the lattice is comprised of two sets of single micro-rings: site rings and link rings. Each site ring hosts the same family of TE resonant modes with discrete frequencies labeled by a mode number $\mu$. The integrated dispersion for the single rings, the effective evanescent coupling rate between two rings, and the hopping phase as a function of mode number are denoted $D_{\rm{int}}(\mu)$, $J(\mu)$, and $\phi(\mu)$, respectively. All these frequency-dependent parameters can be obtained with small-scale 3D FDTD simulations, i.e., for a waveguide and a directional coupler, respectively.

For example, Fig.~\ref{FigSI: 3D FDTD phi}A shows the wavelength-dependence of the hopping phase. Starting from the transfer matrix method~\cite{hafezi2013imaging}, one obtains $\phi=2\pi n_{\rm{eff}}x/\lambda\equiv\beta x$. The displacement of link rings denoted by $x$ is fixed such that for the pump mode $\mu=0$ (where $\lambda_0 = 1550$~nm), the hopping phase is $\phi_0= \beta(\lambda_0)x  =\pi/2$. For $\mu\neq0$ with a wavelength $\lambda_\mu$, the expression for the hopping phase reads:
\begin{equation}
    \phi(\mu)=\phi_0\frac{\beta(\mu)}{\beta(0)}=\phi_0\frac{\lambda_0}{n_{\rm{eff}}(\lambda_0)}\frac{n_{\rm{eff}}(\lambda_\mu)}{\lambda_\mu}.
\end{equation}
\noindent The resulting effective refractive index and hopping phase as a function of frequency are plotted in Fig.~\ref{FigSI: 3D FDTD phi}B.

Moreover, the single-mode tight-binding Hamiltonian~\ref{eqn:single_mode_tb_H_iqh} requires the link rings to be anti-resonant relative to the site rings so that they act as bus waveguides for site ring modes. Mathematically, this means $\beta\eta=(2m+1)\pi$, where m is an integer and $\eta$ is the length difference between a site ring and a link ring.  In particular, the lattice designed in this work is such that $\eta=440$ nm, which satisfies $\beta \eta=\pi$ at 1550~nm. However, this anti-resonant condition cannot be simultaneously satisfied over a large frequency range due to dispersion. To account for the deviation from this condition away from central wavelength of 1550~nm, the Hamiltonian can be corrected using an additional term $-J(\mu)\rm{cot}[\beta(\mu)\eta/2]$ to the on-site potential and a modified coupling strength $J(\mu)/\rm{sin}[\beta(\mu)\eta/2]$, as described in~\cite{hafezi2013imaging}. We note that in our specific case where $m=0$ and $\phi_0=\pi/2$, this results in $\beta(\mu)\eta/2=\phi(\mu)$. Therefore, the corrections in both the on-site potential and the coupling strength can be expressed as functions of the hopping phase $\phi(\mu)$. Finally, such a substitution is only valid when $J(\mu)/|\rm{sin}[\beta(\mu)\eta/2]|\ll D_1$, which is constantly satisfied in this work. 

To calculate $D_{\rm{int}}(\mu)$, we performed comprehensive 3D FDTD simulations of SiN waveguides with a variety of geometries, as shown in Fig.~\ref{FigSI: 3D FDTD WG dispersion}. We obtained the integrated dispersion $D_{\rm{int}}(f)$ of the waveguide with different geometries by numerically sweeping the pump wavelength from 500~nm (600 THz) to 2500~nm (120 THz). We note that higher order dispersion beyond the 4-th order is neglected in this work.

Next, to estimate $J(f)$, we performed simulations of directional couplers, as shown in Fig.~\ref{FigSI: 3D FDTD J}A. We then experimentally estimate~\cite{hafezi2011robust} $J$ for 6 fabricated topological lattices labeled from A1 to C2, as shown in colored scatter plots in Fig.~\ref{FigSI: 3D FDTD J}B. These simulated results show good agreement with experimental estimates at 1550 nm (193 THz). 

Finally, we obtain the mode-dependent $D_{\rm{int}}(\mu)$ and $J(\mu)$ by discretizing $D_{\rm{int}}(f)$ and $J(f)$ using the FSR estimated from linear drop spectrum experiments, while keeping $\mu=0$ at 1550 nm (193 THz).

\begin{figure*}[h]
\centering
\includegraphics[width=0.8\textwidth]{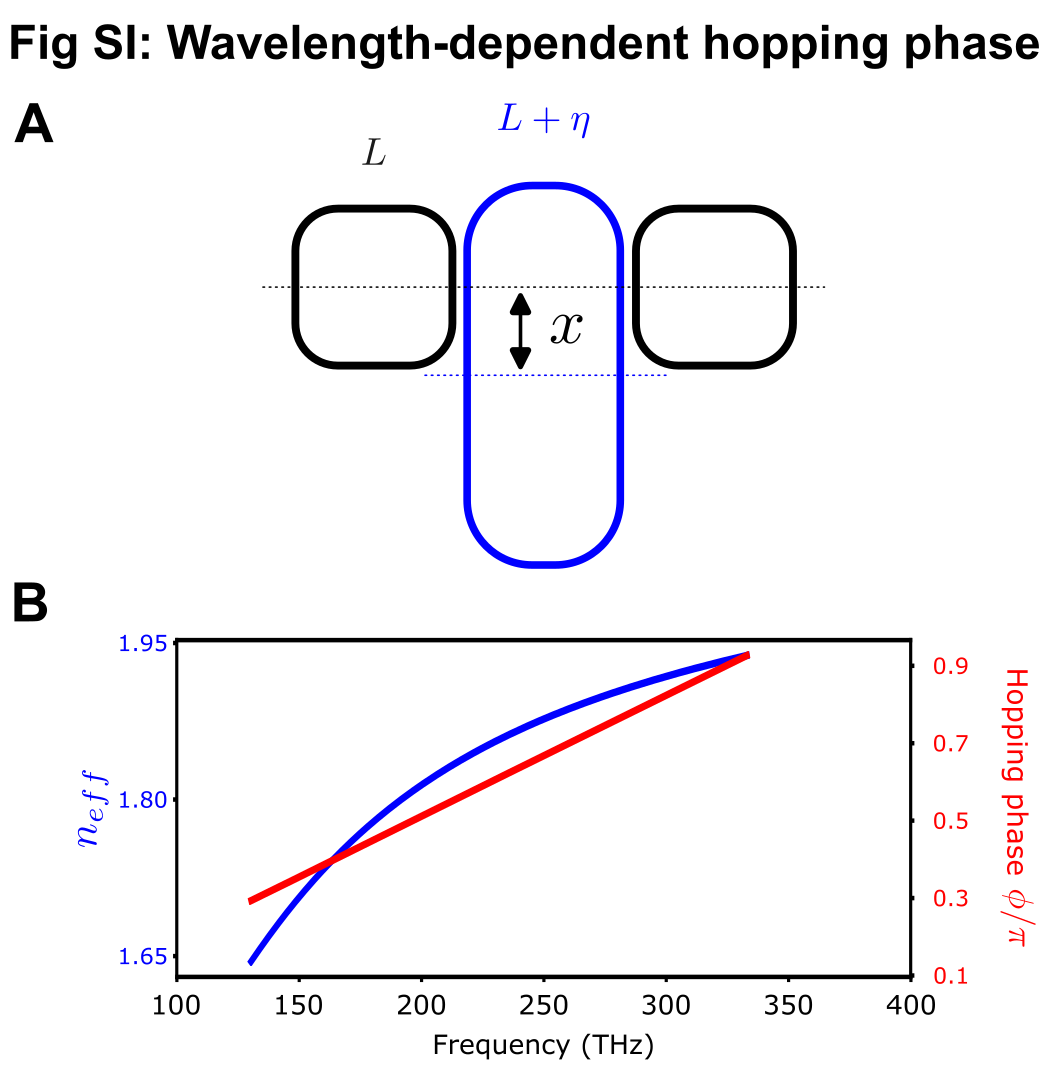}
\caption{\textbf{(A)} Schematic of two-site rings (black) coupled through a link ring (blue). The total lengths of individual site and link rings are $L$ and $L+\eta$, respectively. By shifting the link ring with a deviation $x$, one obtains a hopping phase $\phi=\beta x$, where $\beta=2\pi n{\rm_{eff}}/\lambda$ is the propagation constant. \textbf{(B)} Dispersion of the waveguide that has the same cross-section as the site and link rings (blue), and the frequency-dependent hopping phase (red). }
\label{FigSI: 3D FDTD phi}
\end{figure*}

\begin{figure*}[h]
\centering
\includegraphics[width=0.6\textwidth]{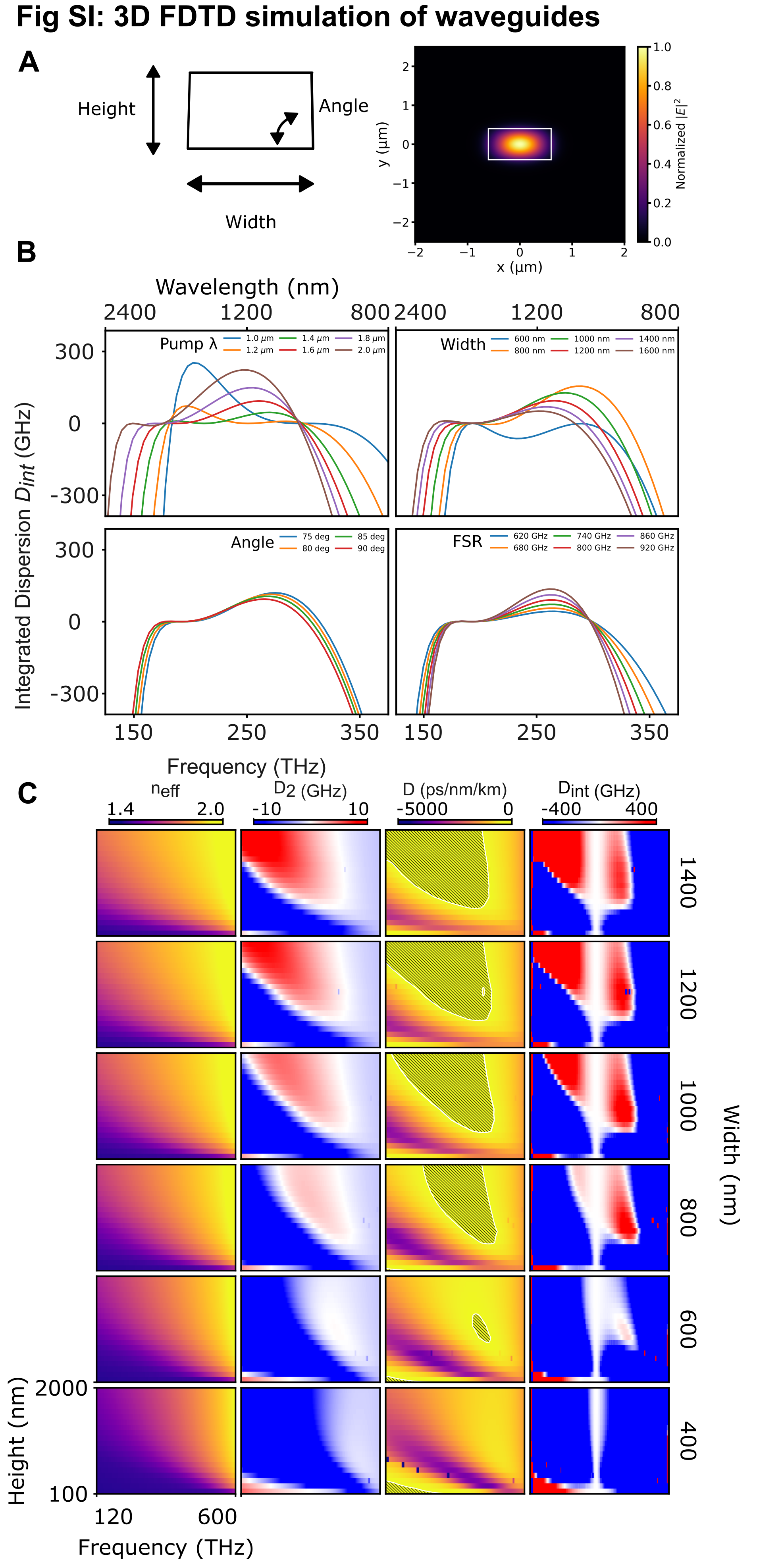}
\caption{\textbf{(A)} Schematic of the waveguide cross section (left) and a representative simulation of the field profile of a TE mode. \textbf{(B)} Integrated dispersion simulation of varying pump wavelength, waveguide width, and waveguide angle for a waveguide with fixed height of 800 nm. The default width, angle, pump wavelength, and the FSR in the four panels are 1200 nm, 90 degrees, 1550 nm, and 774 GHz. \textbf{(C)} Comprehensive simulation of the effective refractive index $n_{\rm{eff}}$, group velocity dispersion (GVD) $D_2$, chromatic dispersion $D$ and integrated dispersion $D_{\rm{int}}$ with varying waveguide height and width. The cross-section of the waveguide is fixed as a rectangle (angle being 90 degrees).}
\label{FigSI: 3D FDTD WG dispersion}
\end{figure*}

\begin{figure*}[h]
\centering
\includegraphics[width=0.6\textwidth]{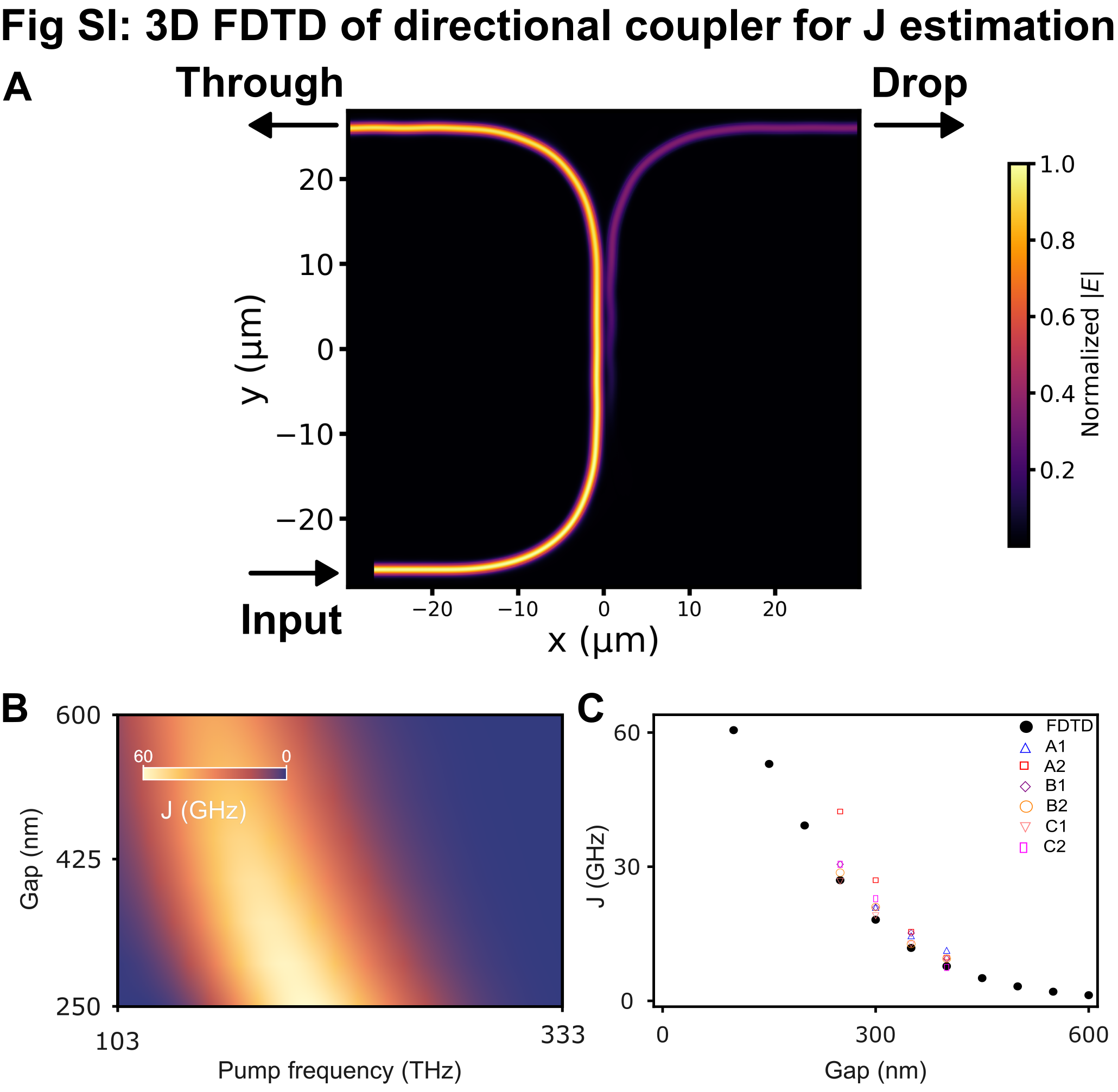}
\caption{\textbf{(A)} A schematic of a simulated directional coupler and its field profile. By varying the gap between the couplers, we control the strength of the evanescent coupling $J$. \textbf{(B)} Simulated sweep of the gap size and pump wavelength dependence for $J$ estimation. \textbf{(C)} For a fixed pump of 1550 nm (193 THz), we simulate $J$ values for different gaps ranging from 100 nm to 600 nm (black), and compare the FDTD simulations against the experimentally estimated results from devices labeled from A1 to C2. }
\label{FigSI: 3D FDTD J}
\end{figure*}
%
\newpage
\section{Two-dimensional integrated dispersion grid and lattice-scale dispersion engineering}\label{sm:2ddispersion}
To highlight the dispersion features of our IQH lattice, we reveal the two-dimensionality of our lattice's integrated dispersion. The integrated dispersion for a single ring describes the deviation of each mode from the equidistant grid separated by the free spectral range (FSR), and centered at the pump mode. Since there is only one timescale within a single ring, its integrated dispersion is a one-dimensional function $D_{\rm{int}}(\mu)$. For a topological lattice, a second scale corresponding to the super-modes of the lattice is introduced. Therefore, the integrated dispersion is promoted from a one-dimensional discretized curve to a two-dimensional grid, which can be modified by the topological model and all the frequency-dependent parameters.

To illustrate the effects of the frequency-dependent parameters on the two dimensional integrated dispersion, we performed 3D FDTD simulations as described in SI section~\ref{sm:dtb}, as shown in Fig.~\ref{FigSI: tight-binding_modification}A. By sequentially adding the topological model and frequency-dependent parameters, one can solve the eigenvalues of the dispersive tight-binding Hamiltonian~\ref{eqn:dispersive_tight_binding} for different single-ring modes $\mu$ and calculate the two-dimensional dispersion grid, as shown in Fig.~\ref{FigSI: tight-binding_modification}B. These results demonstrate the ability of these parameters to tune the dispersion independently. We emphasize that the topological model, the lattice size, the coupling gap size, and the hopping phase are all independent of single-ring geometries. Therefore, these tunabilities are independent of the waveguide dispersion of waveguides.

To further demonstrate the frequency-dependent parameters' tunabilities of the integrated dispersion, we focused on a single parameter $J(\mu)$, whose strength can be controlled with different coupling gaps, as shown in Fig.~\ref{FigSI: gap_tunability}A.  We then calculated the two-dimensional integrated dispersion of the IQH lattices with different gaps, as shown in Fig.~\ref{FigSI: gap_tunability}B. We observed that changing the gap size clearly modifies the topological bandwidth. In particular, a smaller gap typically corresponds to a larger topological bandwidth of the integrated dispersion. 

Next, to experimentally verify the tunability of $J(\mu)$, we designed three IQH lattices that differ only in their gap sizes, performed drop port measurements, and plotted the linear dispersion, as shown in Fig.~\ref{FigSI: gap_tunability_exp}A. Moreover, we compared the experimental results against simulations with the input/output formalism using realistic parameters obtained from 3D FDTD simulations. The extrinsic coupling rate $\kappa_{ex}(\mu)$ also originates from evanescent coupling; therefore, its wavelength dependency is treated in the same way as $J(\mu)$. The intrinsic loss is taken as a constant, and it is estimated from linear drop port experiments near 1550 nm, with an estimated value of $\kappa_{in}\approx$~0.35~GHz. Remarkably, the simulations show great agreement with the experiments, revealing both the tunability of $J(\mu)$ and the clear topological bands. In particular, two chiral edge bands display a spectral shift along different modes, which is caused by the frequency-dependent hopping phase. These results further support the observations in the main text.
\begin{figure*}[h]
\centering
\includegraphics[width=0.99\textwidth]{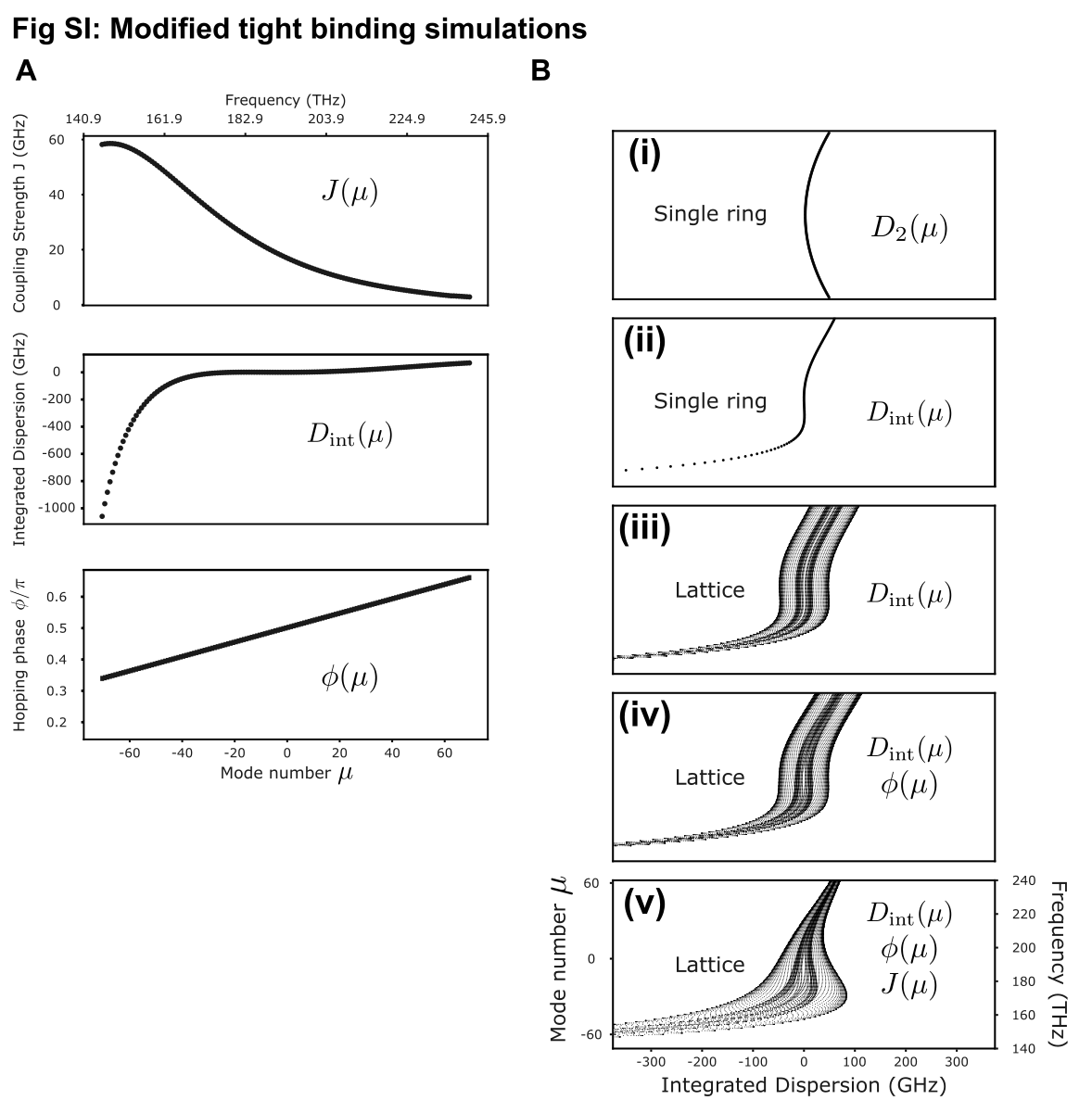}
\caption{\textbf{(A)} Frequency-dependent parameters $J(\mu), D_{\rm{int}}(\mu)$ and $\phi(\mu)$ obtained from 3D FDTD simulations. \textbf{(B)} Simulated integrated dispersion for: (i) single ring with only second-order dispersion $D_2(\mu)$. (ii) A real single ring used in the lattice. Higher orders of dispersion are considered. (iii) IQH lattice, but $J(\mu)$ and $\phi(\mu)$ are constant. The dispersion is promoted to two dimensions with a topological structure. (iv) IQH lattice, but $J(\mu)$ is constant. The introduction of frequency-dependent hopping phase $\phi(\mu)$ modulates the curvature of topological edge modes. (v) Real IQH lattice that considers all frequency-dependent parameters. The frequency-dependent evanescent coupling rate $J(\mu)$ further deforms the integrated dispersion surface.}
\label{FigSI: tight-binding_modification}
\end{figure*}
\begin{figure*}[h]
\centering
\includegraphics[width=0.99\textwidth]{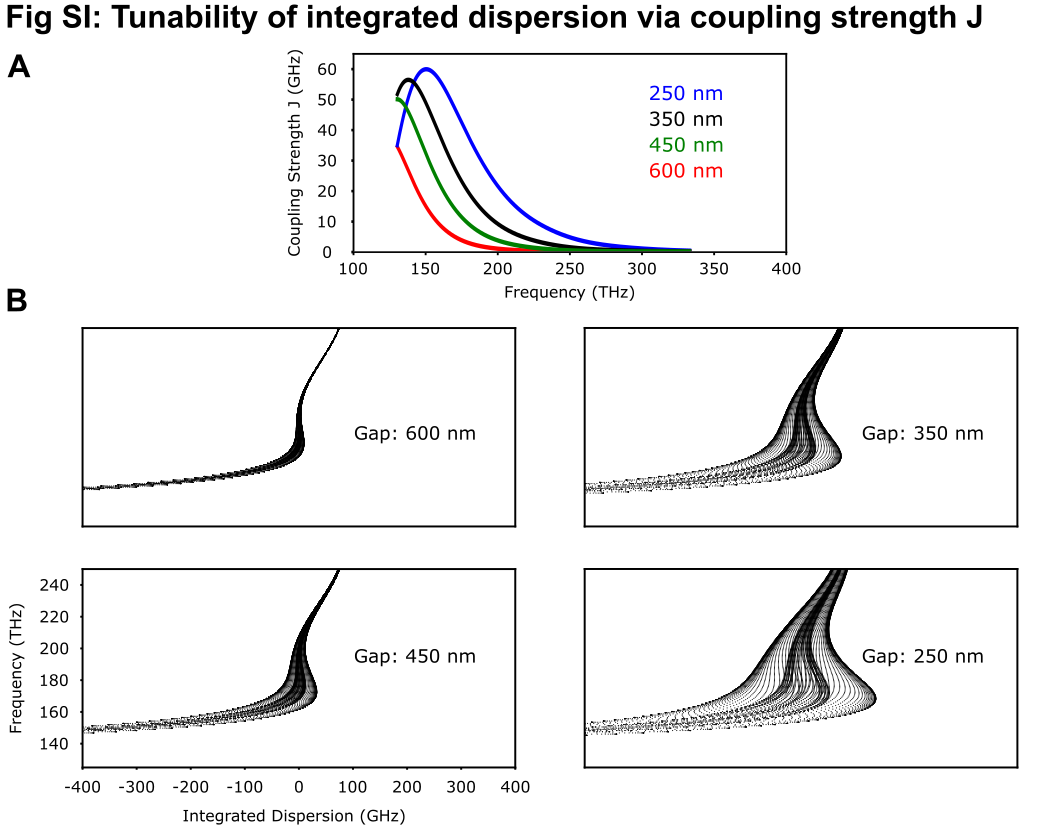}
\caption{\textbf{(A)} Frequency-dependent evanescent coupling rate $J$ as a function of gap size for a directional coupler with the same cross section as the single rings used in this work. \textbf{(B)} Two-dimensional integrated dispersion grid of the IQH lattice with varying gaps.}
\label{FigSI: gap_tunability}
\end{figure*}
\begin{figure*}[h]
\centering
\includegraphics[width=0.99\textwidth]{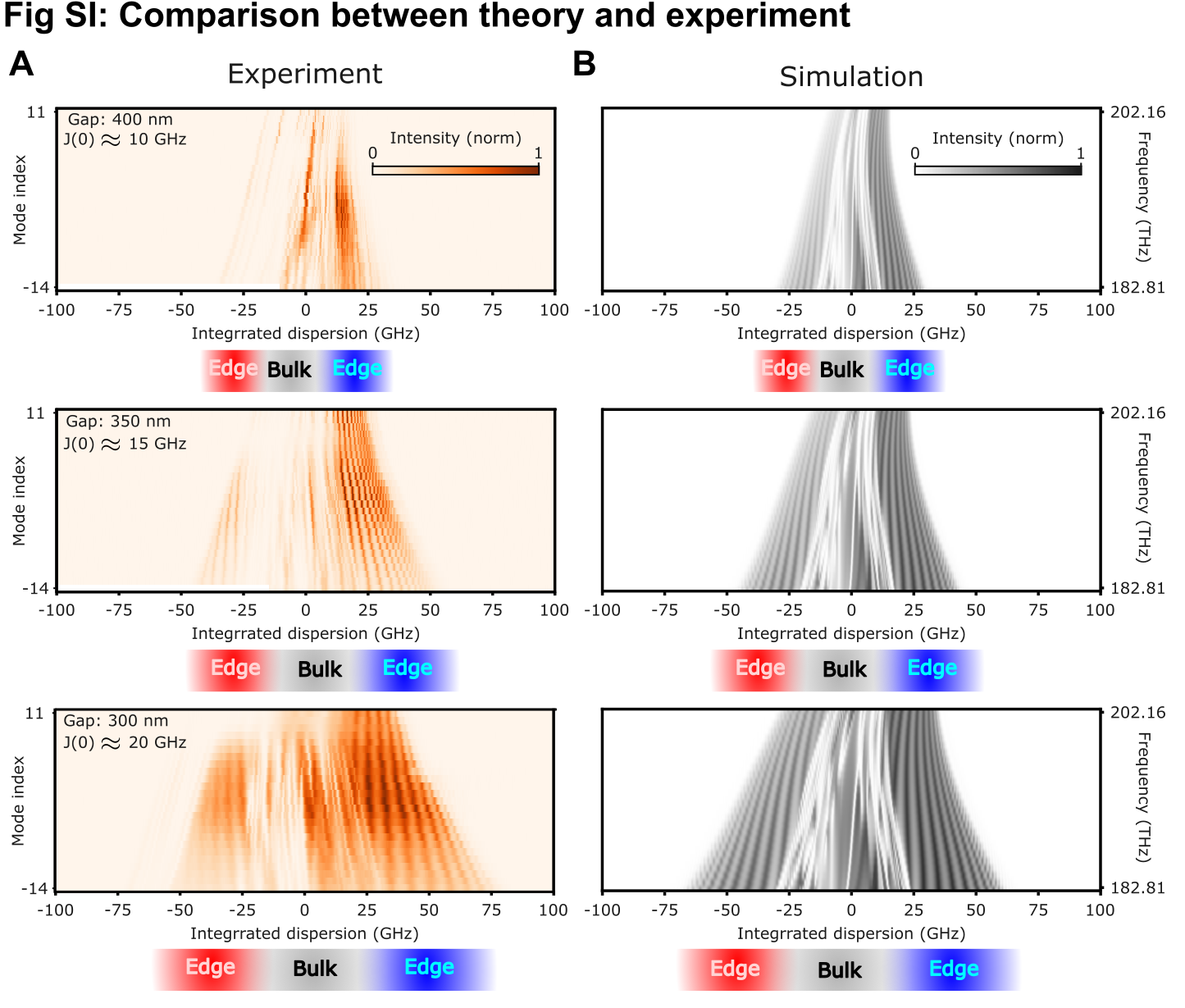}
\caption{The varying AGF landscape for lattices with different gaps. The plots are made with low-power drop spectrum measurements that range from 1480 nm to 1640 nm, covering an estimated number of 25 single-ring modes. \textbf{(A)} Experimentally measured two-dimensional dispersion for IQH lattices with gaps 400 nm, 350 nm and 300 nm. These gaps correspond to an estimated evanescent coupling rate 10 GHz, 15 GHz and 20 GHz at the central mode $\mu=0$ (1550 nm, or 193 THz), respectively. \textbf{(B)} Simulated two-dimensional dispersion that corresponds to the real parameters in (A). } 
\label{FigSI: gap_tunability_exp}
\end{figure*}
\newpage
%
\section{Lugiato-Lefever simulations of multi-modal AGF systems}
\label{sm:lle}
In this section, we discuss the Lugiato-Lefever equations used for topological IQH comb simulations. We assume that all the site rings of the lattice are identical and can be labeled by a Cartesian coordinate ($x,y$). For a square IQH lattice with size $N$, there are $N\times N$ site rings; therefore, $x,y$ goes from 1 to $N$, respectively. The intra-cavity field amplitude of the ring is denoted as $E_{x,y}(\theta,t)$, where $\theta$ is the intra-cavity coordinate. By performing the Fourier transform on $E_{x,y}(\theta,t)$ over $\theta$, one obtains the field amplitude in the mode domain $a_{m}(\mu,t)$. The coupled mode equation for an IQH topological frequency comb is:
\begin{equation}
\begin{aligned}
\frac{d a_{x,y}(\mu)}{d t}
&= \,i\delta\,a_{x,y}(\mu)
   \;+i\Big<[\mathcal{H}_\mathrm{IQH}(\mu)-\omega_0-D_1\mu,\hat{a}_{x,y}(\mu)]\Big> \\
&\quad +\, i\gamma\,\mathcal{FT}\mathbf{\{}
     \bigl|E_{x,y}(\theta)\bigr|^{2} E_{x,y}(\theta)\mathbf{\}}\,  \\
&\quad -\,\bigl(\kappa_{\mathrm{ex,\mu}}\,\delta_{\mathrm{IO}} + \kappa_{\mathrm{in}}\bigr)\,a_{x,y}(\mu)
     \;+\; \delta_{\mathrm{IO}}\delta_{\mu,0}\,\sqrt{2\kappa_{\mathrm{ex,\mu}}}\mathcal{E}\,.
\end{aligned}
\label{eqn:cmt_iqh_si}
\end{equation}
\noindent where $D_1$ is the free spectral range (FSR) of a site ring, $\omega_p$ is the pump frequency, and $\omega_0$ is the resonant frequency of the $\mu=0$ mode for the site ring. The nonlinear coefficient $\gamma$ is given by:
\begin{equation}
    \gamma=\frac{c\omega_0n_2}{n_0^2A_{eff}L}.
\end{equation}
\noindent where $\omega_0$ is the pump mode frequency, $n_0 (n_2)$ is the linear (nonlinear) refractive index, $A_{eff}$ is the effective mode area for a single ring mode, and $L$ is the cavity length. Here, $\gamma$ is taken as a constant.
The intrinsic loss rate for a site ring is denoted as $\kappa_{\mathrm{in}}$, which is assumed to be a constant value. The extrinsic coupling rate $\kappa_{\mathrm{ex,\mu}}$ depends on the mode number $\mu$, as it physically originates from the evanescent coupling between the bus waveguide and the input/output ring. The pump is assumed to be monochromatic, and its field amplitude is denoted as $\mathcal{E}$. Finally, we define the detuning as $\delta\equiv\omega_p-\omega_0$ and normalize the equation with the following conventions:
\begin{align*}
\frac{\omega_{0}}{J_0} &\rightarrow \omega_{0}, &
\frac{\omega_{p}}{J_0} &\rightarrow \omega_{p}, &
\frac{D_{1}}{J_0} &\rightarrow D_1, &
\frac{\kappa_{\mathrm{ex,\mu}}}{J_0} &\rightarrow \kappa_{\mathrm{ex,\mu}}, &
\frac{\kappa_{\mathrm{in}}}{J_0} &\rightarrow \kappa_{\mathrm{in}} \\[6pt]
J_0 t &\rightarrow t, &
\sqrt{\tfrac{\gamma}{J_0}}\,a_{m,\mu} &\rightarrow a_{m,\mu}, &
\sqrt{\tfrac{2\kappa_{\mathrm{ex,\mu}}\gamma}{J_0^{3}}}\,\mathcal{E} &\rightarrow \mathcal{F},  & \frac{\mathcal{H}_{m,\mu}}{J_0}\rightarrow \mathcal{H}_{m,\mu},
\end{align*}
\noindent where $J_0$ is the evanescent coupling rate at the pump wavelength, which corresponds to $\mu=0$. We note that, with such a normalization convention, the simulation is performed with dimensionless parameters, which is more convenient.
%
%
\section{Linear drop port measurement and simulation of IQH lattices}
\label{sm:linear}
To better estimate the experimental parameters and demonstrate the topological properties of the IQH lattices, we performed drop port transmission and group delay experiments, as well as simulations, for two IQH lattices with different $J$ values. In particular, we swept over the topological band structure near 1548 nm (193.66 THz), where $J$ values for these lattices are approximately 20 GHz and 15 GHz, respectively. The results are shown in Fig.~\ref{FigSI: Transmission and group delay}. With the estimated parameters, where the intrinsic loss rate $\kappa_{in}=0.35$~GHz and the extrinsic loss rate $\kappa_{ex}=3$~GHz, the simulations show very good agreement with the experiments. 

To further demonstrate the transport properties of the topological edge bands, we performed group delay spectrum analysis, as shown in the third row of each panel in Fig.~\ref{FigSI: Transmission and group delay}. We observed that, due to the locations of the input and output rings, the two chiral edge states travel a path of one lattice side length (short edge) or three lattice side lengths (long edge). Therefore, the difference between the delay times for the two chiral edge bands corresponds to half of the edge mode's round-trip time, which is measured as 0.22 ns for $J\approx 20$ GHz and 0.32 ns for $J\approx 15$ GHz. This matches our expectation, where lattices with larger $J$ correspond to faster transport along the edge of the lattice. We note that the exact round-trip time depends on the spectral location of the pump frequency inside the edge bands, which is proportional to but not linearly dependent on $1/J$.  
\begin{figure*}[h]
\centering
\includegraphics[width=0.99\textwidth]{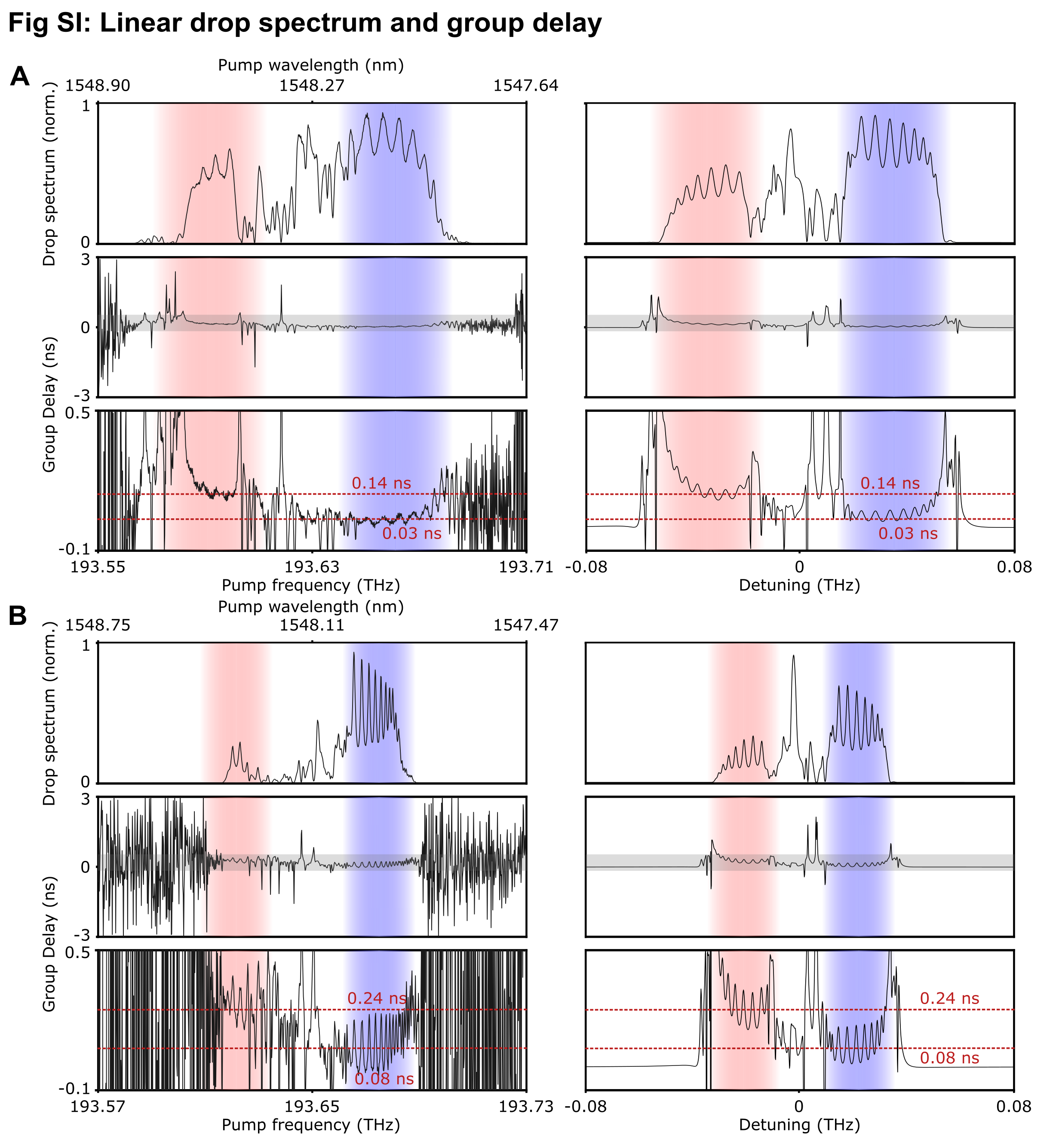}
\caption{\textbf{(A)} Left: the measured drop spectrum (first row) and delay spectrum (second and third row) of an IQH lattice with $J\approx 20$ GHz at $\mu=0$. The third row is the zoomed-in plot of the grey-shaded area in the second row. Right: the corresponding simulation with realistic parameters. Two chiral edge bands are highlighted with red and blue shades, within which we label the measured average delay time for both chiral edge bands in the third row. \textbf{(B)} Same plots as (A) for another IQH lattice with $J\approx 15$ GHz at $\mu=0$. }
\label{FigSI: Transmission and group delay}
\end{figure*}
\newpage
\section{Constructing the integrated dispersion plot}\label{sm:dint_heatmap}

To illustrate how the integrated dispersion of the lattice is reconstructed from drop port measurements, we performed a detailed technical analysis of the multi-shot realization of AGFs, as shown in Fig.~\ref{FigSI: dint_heatmap}. We begin by sweeping the pump frequency across multiple modes separated by an FSR of $\approx$~773.88~GHz, a number estimated from the drop port spectrum of a site ring that comprises the lattice. Each mode contains a topological structure with a bandwidth of $\approx$~200~GHz. Next, we created a waterfall plot by vertically stacking the drop spectrum of each mode $\mu$ separated by an FSR. Finally, such a waterfall plot is re-plotted as a heatmap, where colors represent the transmission. We note that this method of obtaining the integrated dispersion of a lattice also works for the single-shot realization of AGFs.

\begin{figure*}[h]
\centering
\includegraphics[width=0.8\textwidth]{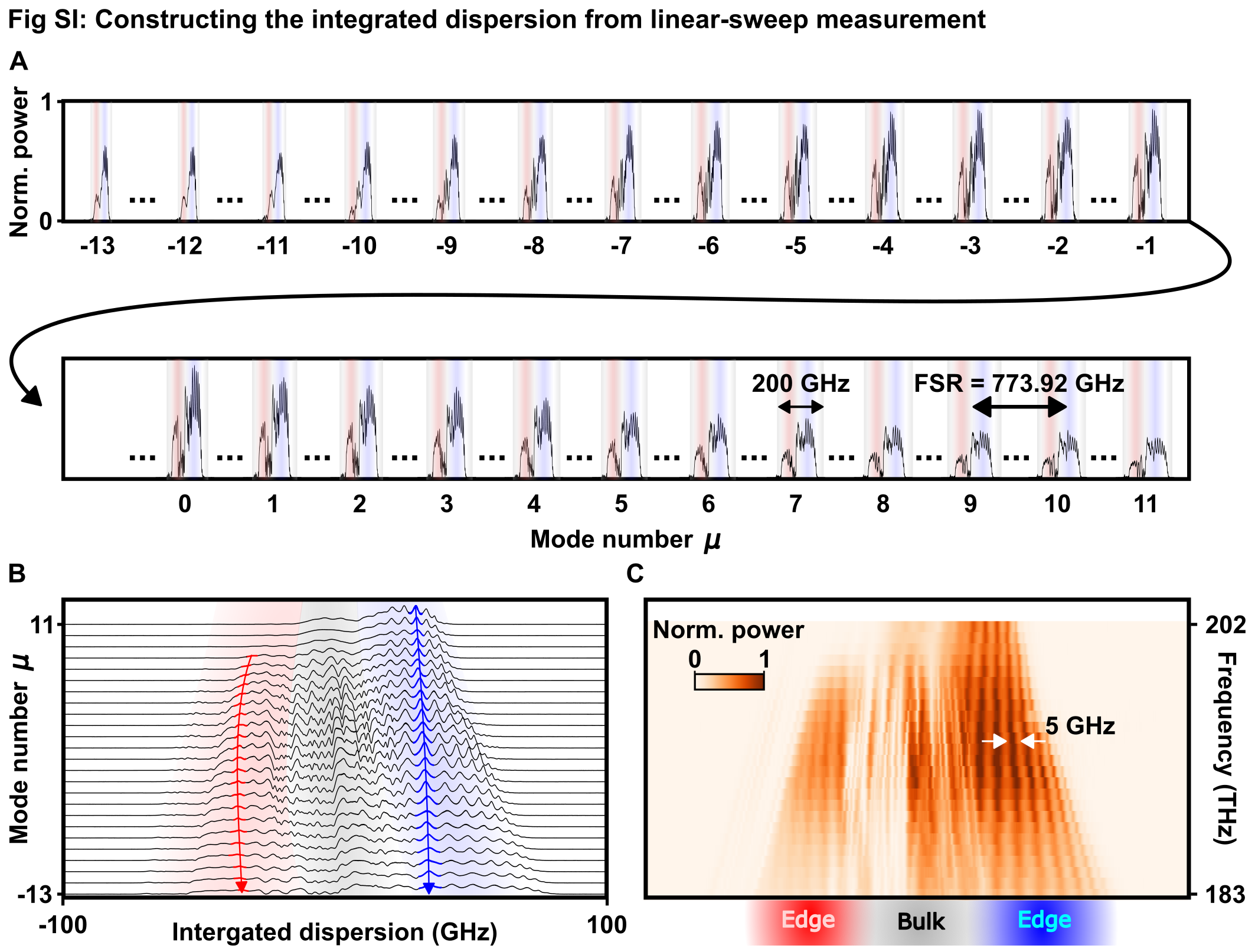}
\caption{\textbf{(A)} Representative linear-frequency sweep spectra showing successive cavity resonances, which are unwrapped and indexed by the integer mode number $\mu$. The local frequency window (200~GHz) and the global free spectral range (FSR $\approx$~773.88~GHz) establish the mapping between measured resonance frequencies and discrete cavity modes.
\textbf{(B)} Integrated dispersion $D_{\mathrm{int}}(\mu)$ extracted by subtracting the linear reference $\omega_\mu = \omega_0 + D_1 \mu$ from the measured resonance frequencies. Distinct dispersion trends associated with different mode families are highlighted in red or blue, revealing systematic deviations from uniform mode spacing.
\textbf{(C)} Two-dimensional map of normalized resonance power versus mode number, illustrating the reconstructed multi-mode dispersion landscape. The spectral spacing between two adjacent edge modes is $\approx 5$~GHz.}

\label{FigSI: dint_heatmap}
\end{figure*}
\newpage


\section{Chip-scale 3D FDTD simulations of the drop spectrum and field profile of IQH lattices}\label{sm:chipscale}
To better understand the field profile of the IQH lattices, we performed a chip-scale linear 3D FDTD simulation of single rings and IQH lattices with different sizes and gaps. Since a full-scale 3D FDTD simulation of the real device is time-consuming and power-intensive, we performed the narrow-band simulation by sweeping the pump wavelength across the topological edge band for just one fixed single-ring mode $\mu=0$ near 1550 nm, with a relatively large sweep step of 0.1 nm. The simulated drop spectra for different lattices are shown in the first column of Fig.~\ref{FigSI: 3D FDTD field profile}. Since the spectral spacing of two adjacent edge modes is approximately 20 pm, a simulation step of 0.1 nm is insufficient to resolve individual edge modes. However, by monitoring the field profile at different wavelengths, we observed and identified the two chiral topological edge states, as well as the bulk state, as clearly shown in Fig.~\ref{FigSI: 3D FDTD field profile}. 

To experimentally verify the field profile of the edge states, we destructively pumped the lattice in the edge band with ultra-high power and burned the device, as shown in Fig.~\ref{FigSI: 3D FDTD field profile burn}. Since the lattice is coupled with two bus waveguides, we performed the first experiment by sending light into one of the bus waveguides. In the first experiment, we only burned the input ring, essentially creating a single-ring defect inside the lattice. We then performed the second experiment by sending light into the other bus waveguide and observed a meandering burning path along the edge. We make two remarkable observations. First, the burning path matches the highest intensity path in our 3D FDTD simulations, as shown in Fig.~\ref{FigSI: 3D FDTD field profile burn}B. Second, the second burning path goes around the first burned ring, validating the robustness of the edge state in the nonlinear regime.
\begin{figure*}[h]
\centering
\includegraphics[width=0.99\textwidth]{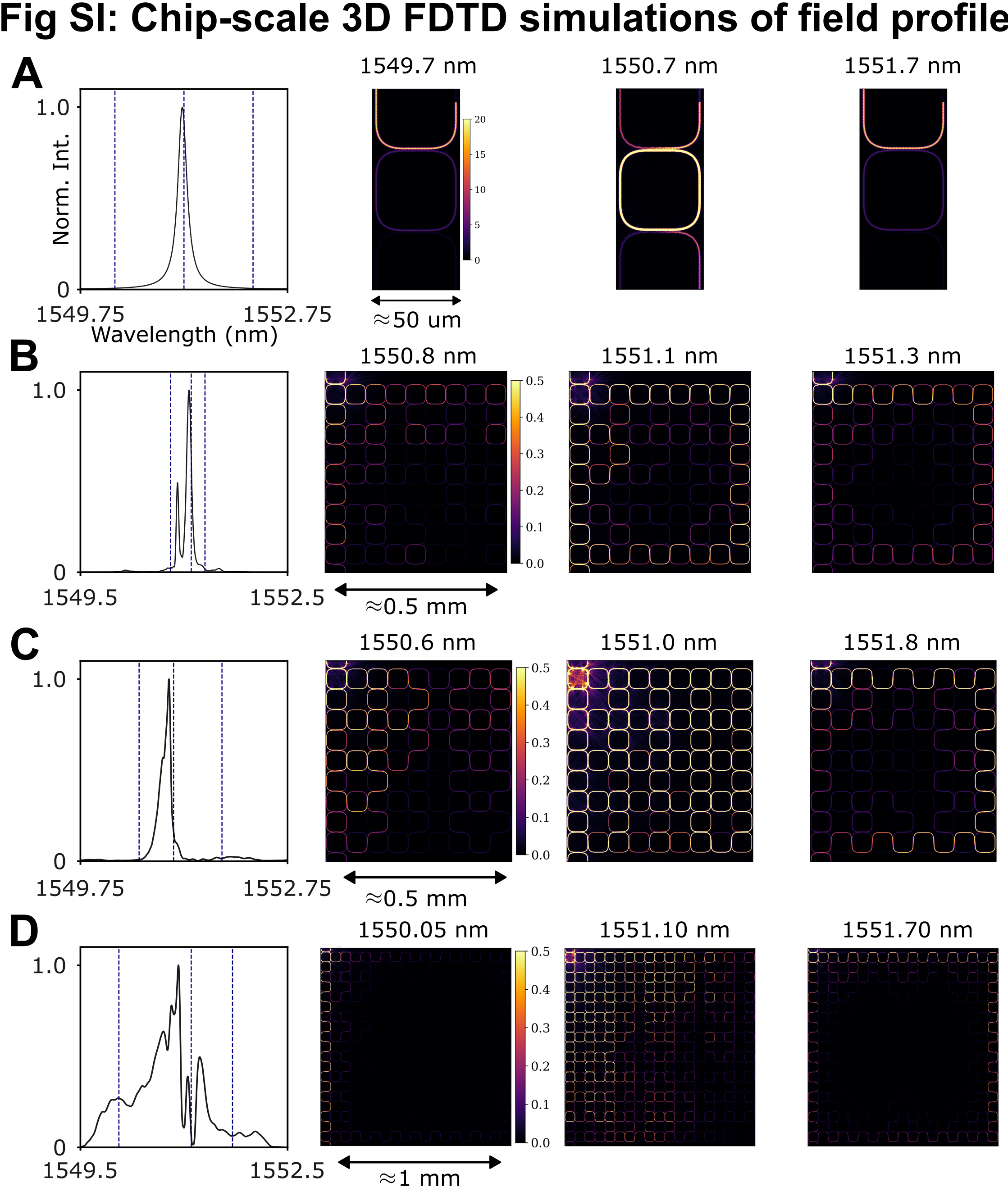}
\caption{Chip-scale 3D FDTD simulations of drop port transmission and field profiles for \textbf{(A)} a single ring resonator, \textbf{(B)} a 5 \texttimes{} 5 IQH lattice with ring gap = 400 nm, \textbf{(C)} a 5 \texttimes{} 5 IQH lattice with ring gap = 300 nm, and \textbf{(D)} a 10 \texttimes{} 10 IQH lattice with ring gap = 300 nm. In each row, the three field profiles correspond to the pump detunings marked by dashed lines in the drop spectrum in the first panel. The simulated single rings (including site and link rings) are identical to our real devices.}
\label{FigSI: 3D FDTD field profile}
\end{figure*}
\begin{figure*}[h]
\centering
\includegraphics[width=0.99\textwidth]{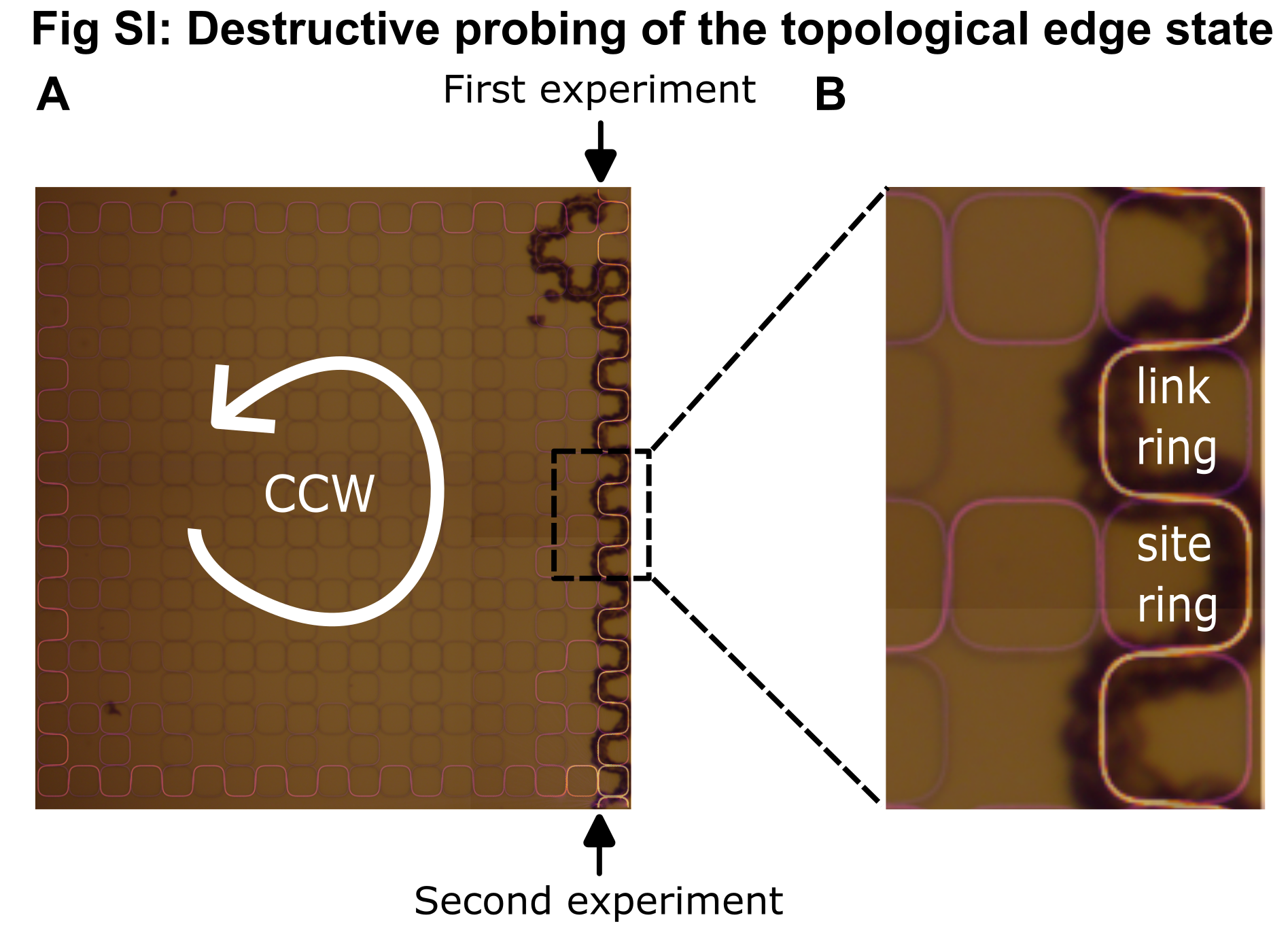}
\caption{\textbf{(A)} Real image of a burned device overlaid with a representative 3D FDTD simulation of the field profile in the same chiral edge band. Two destructive experiments by pumping the edge band with ultra-high power (peak power as high as 700 W) were performed sequentially on both sides of the bus waveguides. \textbf{(B)} A zoomed-in analysis showing the burned edge state with link rings and site rings.}
\label{FigSI: 3D FDTD field profile burn}
\end{figure*}
\newpage
\section{Pump power dependence of comb spectra and integrated dispersion}\label{sm:power}
To study the pump power dependence of the generated IQH combs and their corresponding integrated dispersion, we selected the 300 nm gap lattice presented in the main Fig.~\ref{Fig:generate_comb}B, and sequentially increased the pump power by fixing the pump frequency at 193.67 THz. The generated combs and the integrated dispersion are shown in Fig.~\ref{FigSI: IQH comb power sweep}. We observed that, as pump power increases, the comb broadens, and the integrated dispersion plots on the right panels of Fig.~\ref{FigSI: IQH comb power sweep} extend further to both higher and lower frequencies, approaching an octave.
\begin{figure*}[h]
\centering
\includegraphics[width=0.99\textwidth]{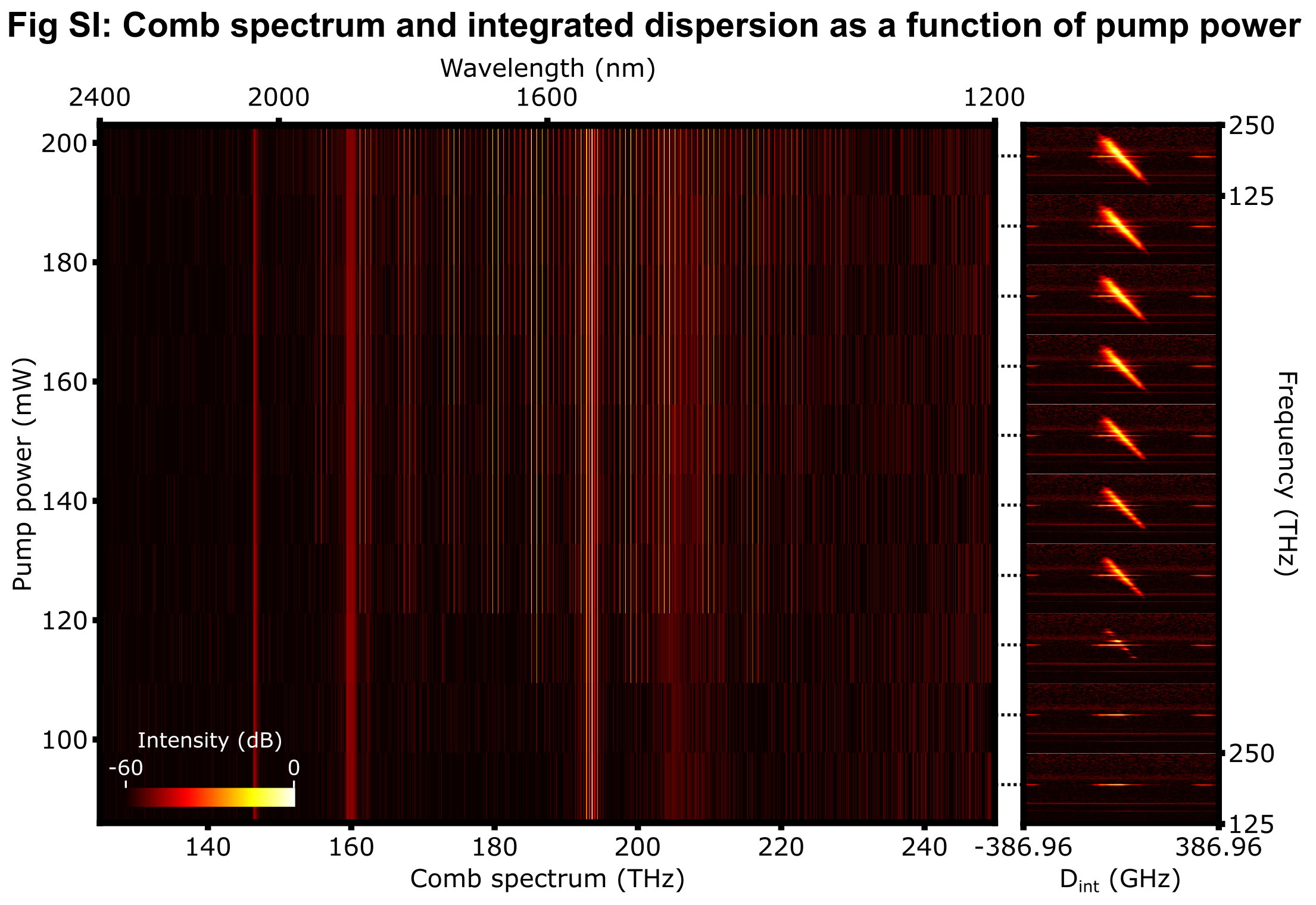}
\caption{For an IQH lattice with a gap of 300 nm, whose coupling strength $J\approx 20$ GHz near 193 THz ($\mu=0$), we fixed the pump freqeuncy at 193.67 THz and performed a pump power sweep. The heatmap of comb spectra is plotted on the left, and the integrated dispersion is plotted on the right.}
\label{FigSI: IQH comb power sweep}
\end{figure*}
\newpage

%
\section{Spatial imaging of generated harmonics}\label{sm:image}

Recently, the concept of nested frequency-phase matching (FPM) for integrated nonlinear photonics was demonstrated on an AQH topological lattice~\cite{mehrabad2025multi}. To demonstrate the nested FPM on IQH lattices, we performed unfiltered harmonic imaging experiments on IQH lattices by collecting the out-of-plane scattered light above the chip, in the same fashion as described in~\cite{mehrabad2025multi}. Remarkably, we observed the second (red), third (green), and fourth (blue) harmonic light, as shown in Fig.~\ref{FigSI: visible imaging}. Interestingly, we observed a single IQH plaquette with green light, where harmonic light is spatially located inside site rings, as shown in Fig.~\ref{FigSI: visible imaging}B. Moreover, we observed in Fig.~\ref{FigSI: visible imaging}D that the harmonic generation is confined to the edge when we pump the edge band, whereas no such confinement is observed when the bulk modes are pumped. The reproduction of higher harmonic generations in a nested system with a different topology further verifies the nested FPM concepts reported in~\cite{mehrabad2025multi}. 
\begin{figure*}[h]
\centering
\includegraphics[width=0.99\textwidth]{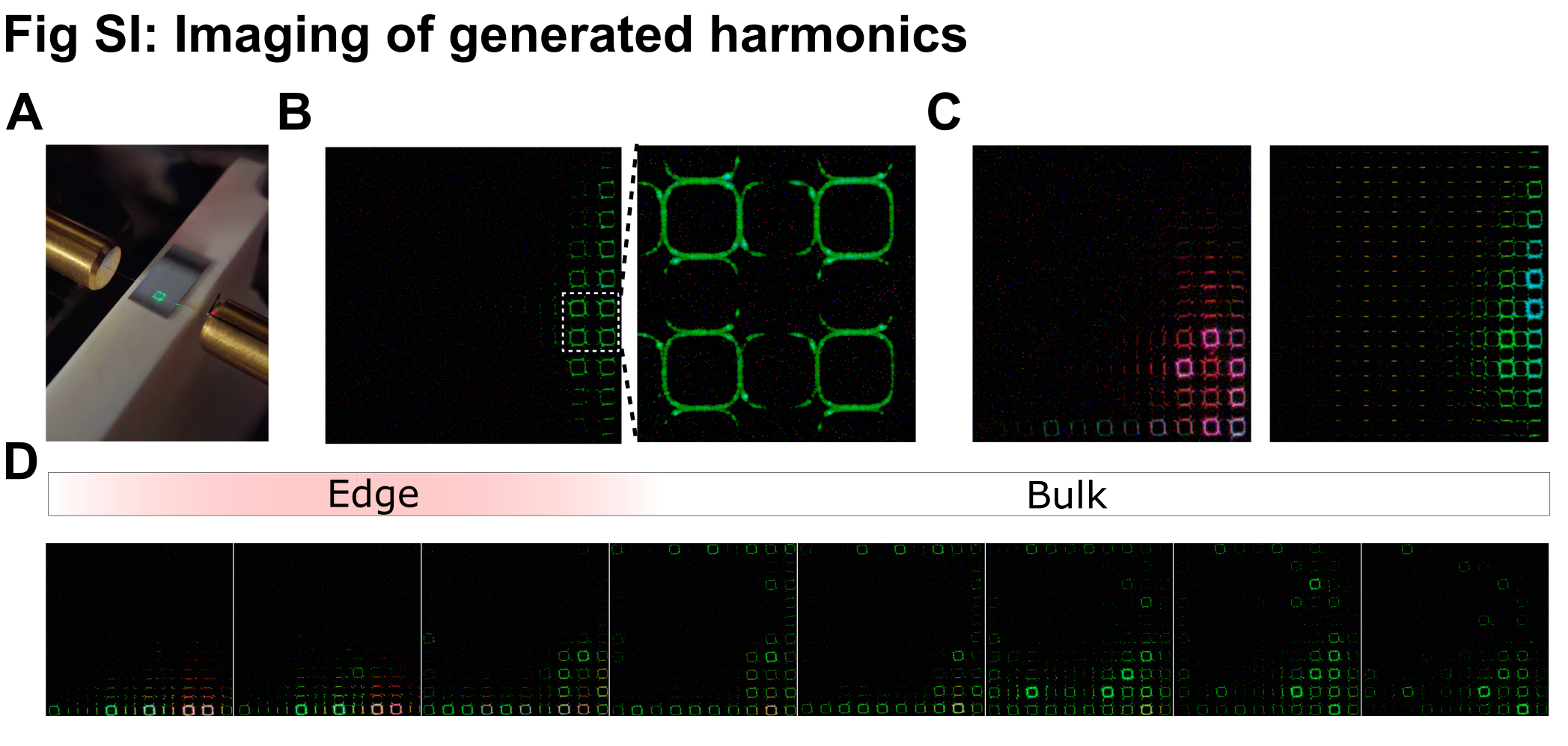}
\caption{\textbf{(A)} Real photo of the chip. By pumping the lattice near 1548 nm (193.66 THz), we generate different harmonics of light that correspond to \textbf{(B)} green (third-harmonic), \textbf{(C)} red (second-harmonic), and blue (fourth-harmonic) light. A zoomed-in image of an IQH plaquette is shown in \textbf{(B)}. \textbf{(D)} Pump-wavelength dependent spatial imaging of the generated third-order harmonics that covers one chiral edge band (colored red) and the bulk band (colored white).}
\label{FigSI: visible imaging}
\end{figure*}
%
\section{Comb generation and integrated dispersion for a single ring resonator}\label{sm:singlering}
To compare the integrated dispersion of our IQH lattice with that of single micro-resonators, we generated combs by pumping a single ring resonator that has the same geometry as the site rings inside the IQH lattice. Fig.~\ref{FigSI: single ring comb dispersion}A shows a representative single ring comb. We observed that while the integrated dispersion of the single-ring comb deviates from the linear dispersion due to nonlinearity, its spectral shape, as well as its curvature, closely follows the linear integrated dispersion. Compared to lattice dispersion, the bandwidth of the single ring's dispersion (along the horizontal axis) is narrower due to the absence of $J$ and lacks signatures of topological bands. We further analyzed the pump wavelength dependence of the comb spectra and the integrated dispersion for the single ring device by sweeping the pump wavelength at a fixed power of 107 mW, as shown in Fig.~\ref{FigSI: single ring comb dispersion}B. We observed that, unlike the topological IQH comb, whose integrated dispersion can be engineered with $J$ and selective pumping within the topological band, the integrated dispersion curve of a single ring across different wavelengths within one resonance remains unchanged. 
\begin{figure*}[h]
\centering
\includegraphics[width=0.99\textwidth]{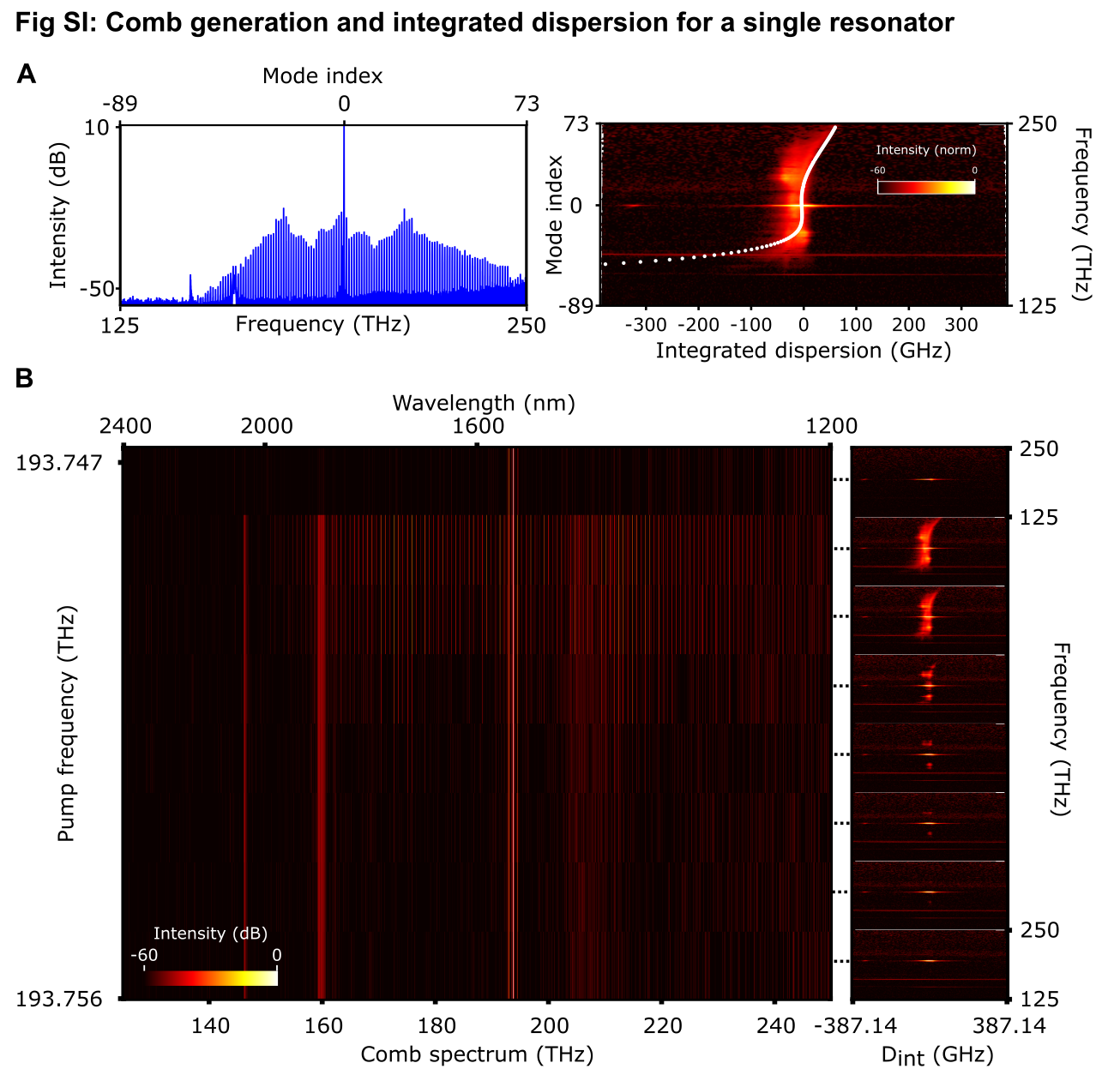}
\caption{\textbf{(A)} Left: a representative comb spectrum generated from a single microresonator. The pump frequency is 193.75 THz, and the pump power is 107 mW. Right: the integrated dispersion of that comb. The white scattered plot represents the linear integrated dispersion obtained from 3D FDTD simulation \textbf{(B)} Comb generation for the same microresonator by fixing the pump power at 107 mW while sweeping the pump frequency across the resonance.}
\label{FigSI: single ring comb dispersion}
\end{figure*}
%
\section{Full pump frequency sweep simulation of IQH comb dispersion}\label{sm:fullsweep}
To demonstrate the tunability of the integrated dispersion of the IQH lattices by pumping different bands of the lattice, we performed nonlinear simulations of IQH combs with realistic parameters by conducting a pump frequency sweep with a fixed pump power on a 300 nm gap lattice, as shown in
Fig.~\ref{FigSI:detuning_sweep}. We observed that the slope of the integrated dispersion changes drastically as the pump sweeps across the topological band, and it locally follows the linear integrated dispersion predicted by the dispersive tight-binding model. These observations suggest that the coherence, as well as the repetition rates of the generated combs, can be modified simply by pumping different frequencies within the topological band.

\begin{figure*}[h]
\centering
\includegraphics[width=0.8\textwidth]{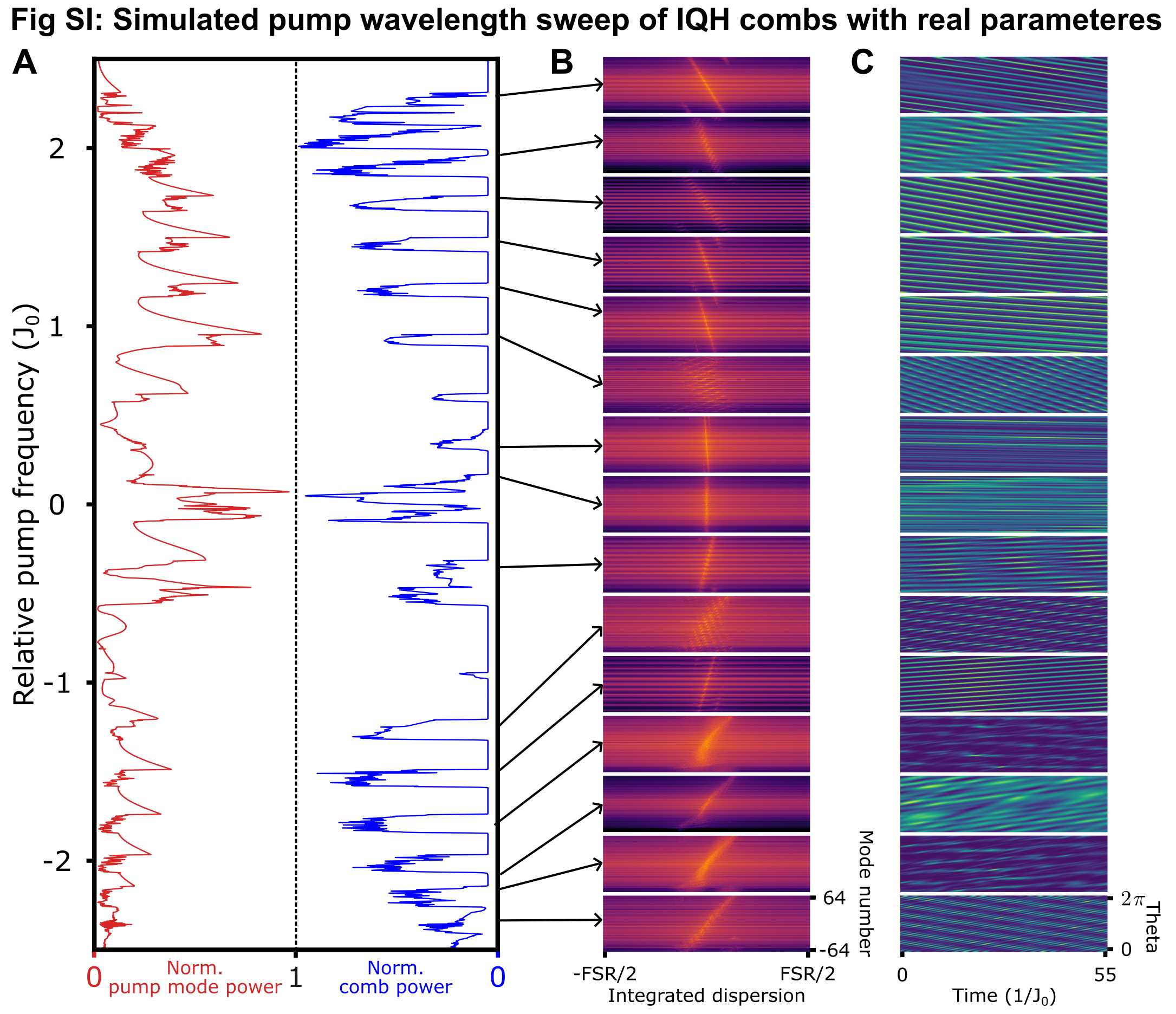}
\caption{\textbf{(A)} For the realistic parameters, we sweep the pump detuning from 2.5~$J_0$ to -2.5~$J_0$ and plot the pump mode power (red) as well as the integrated comb power (blue). \textbf{(B)} The corresponding integrated dispersion plots for different pump detuning inside the output ring. \textbf{(C)} The corresponding spatiotemporal plots of the output ring for different pump detuning.}
\label{FigSI:detuning_sweep}
\end{figure*}
\newpage
\section{High resolution optical spectrum measurements}\label{sm:apex}
To better resolve the nested structure of the IQH combs, we generated an IQH comb by pumping the CW band of the 300~nm~gap lattice and measured the comb with the high-resolution OSA, as shown in Fig.~\ref{FigSI:nesting}A. Moreover, we performed both  linear and nonlinear simulations using realistic parameters, as shown in Fig.~\ref{FigSI:nesting}B. We observed not only a pronounced agreement between simulations and experiments but also clear nested structures of the comb teeth with a characteristic spacing of 5 GHz. 

Moreover, we compared the integrated dispersion of the generated comb against the linear dispersion obtained from multi-shot experiments and simulations, as shown in Fig.~\ref{FigSI:nesting}C,D. We observe that, while pumping one of the chiral edge bands (which here is the CW band), no optical signal is generated on the other chiral edge band (CCW band). This again indicates that the topological robustness against back-scattering is valid under strong nonlinearity and is reproducible using the dispersive tight-binding Hamiltonians.

Next, we performed comprehensive high-resolution optical spectrum measurements of the IQH frequency combs on lattices with 300~nm and 400~nm~gaps, as shown in Figs.~\ref{FigSI:apex_300nm_sweep} and~\ref{FigSI:apex_400nm_sweep}. Across multiple AGFs realized in the nonlinear regime using a single-shot scheme, the IQH combs consistently exhibit fine spectral substructures. Notably, the 300~nm-gap device shows nested features with a characteristic 5~GHz spacing when the pump is swept across the entire edge band.  

\begin{figure*}[h]
\centering
\includegraphics[width=1\textwidth]{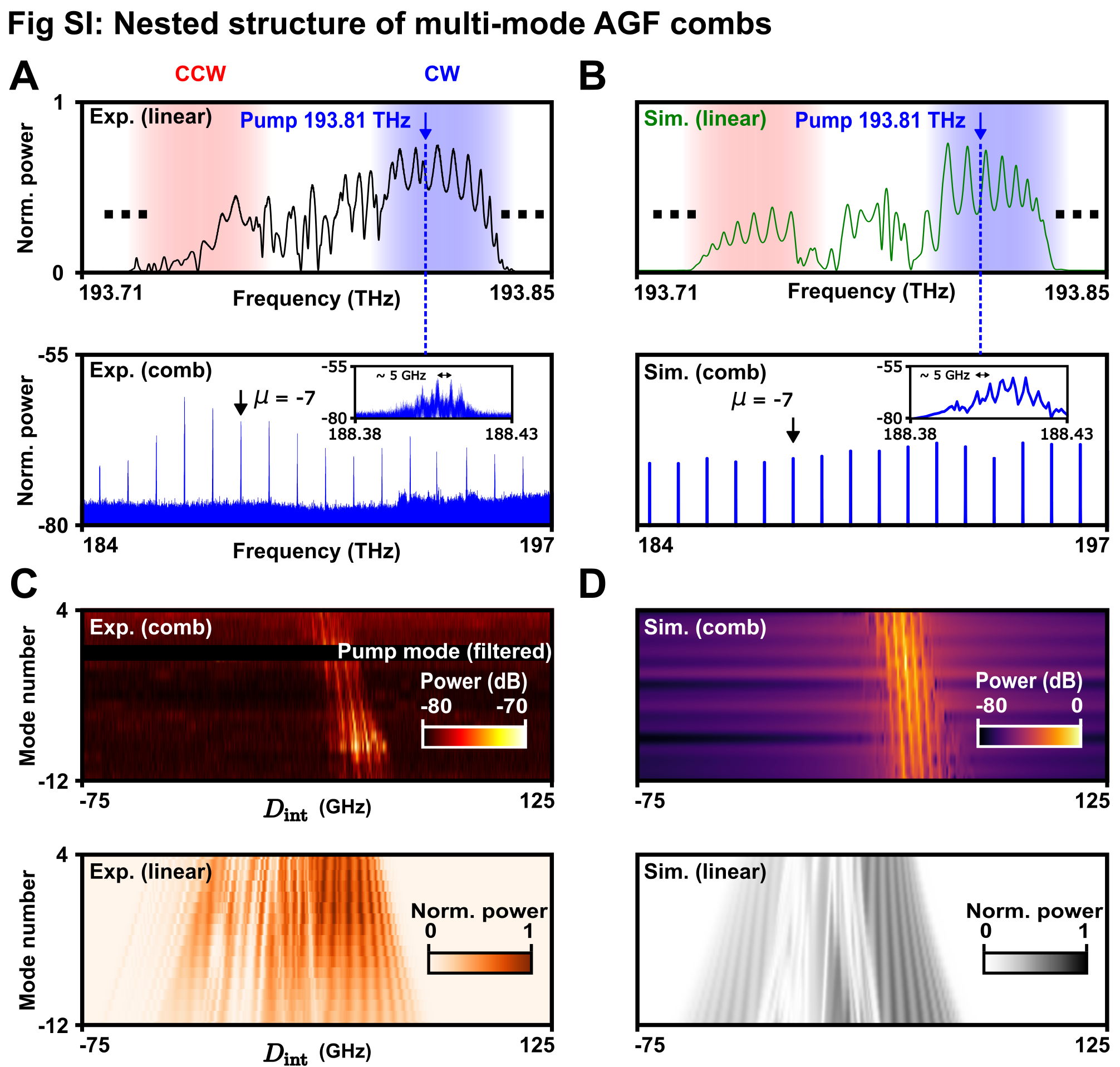}
\caption{\textbf{(A)} Narrow-band drop spectrum of an IQH lattice near 193 THz, corresponding to the mode $\mu$ = 0. This lattice has a ring-to-ring gap of 300 nm, where the CW/CCW chiral edge bands are colored blue and red. By pumping in the CW edge band with 193.81 THz, an IQH frequency comb is generated. After filtering out the pump, the comb is sent to a high-resolution and narrow-band OSA. The inset shows the zoomed-in plot of the comb teeth that correspond to the mode $\mu=-7$, revealing a nested structure of an estimated spacing of $\approx5$ GHz. \textbf{(B)} The exact linear and nonlinear simulations of panel (A). \textbf{(C)} First row: the integrated dispersion plot of the IQH comb generated in panel (A), covering modes $\mu$ from -12 to 4. Second row: The linear integrated dispersion of the same lattice measured within the same frequency range. \textbf{(D)} The exact linear and nonlinear simulations of panel (C).}
\label{FigSI:nesting}
\end{figure*}

\begin{figure*}[h]
\centering
\includegraphics[width=1\textwidth]{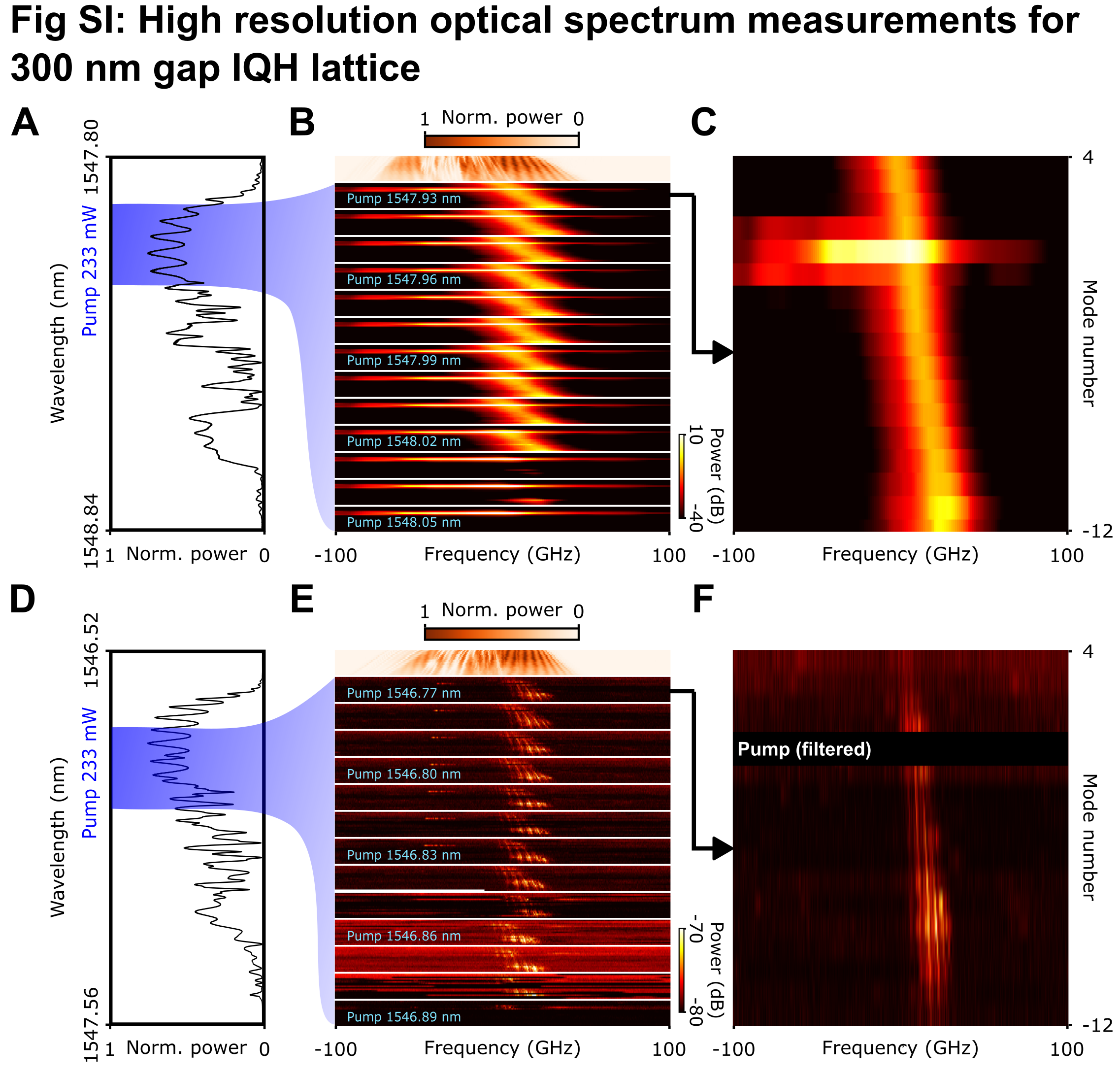}
\caption{\textbf{(A)} The linear drop spectrum of a 300 nm gap IQH device. We generate frequency combs by sweeping the pump wavelength across one of the chiral edge bands with a fixed pump power of 233 mW, as indicated in blue shades. \textbf{(B)} Orange: the linear integrated dispersion plot of the device. Red: the integrated dispersion of the generated IQH combs measured with a 20 pm low-resolution optical spectrum analyzer. \textbf{(C)} Zoomed-in plot of the comb for pump wavelength 1547.93 nm. \textbf{(D-F)} Same plots as panels A-C for a copy of the same lattice, where the combs are measured with a 40 fm high-resolution optical spectrum analyzer.}
\label{FigSI:apex_300nm_sweep}
\end{figure*}
\begin{figure*}[h]
\centering
\includegraphics[width=1\textwidth]{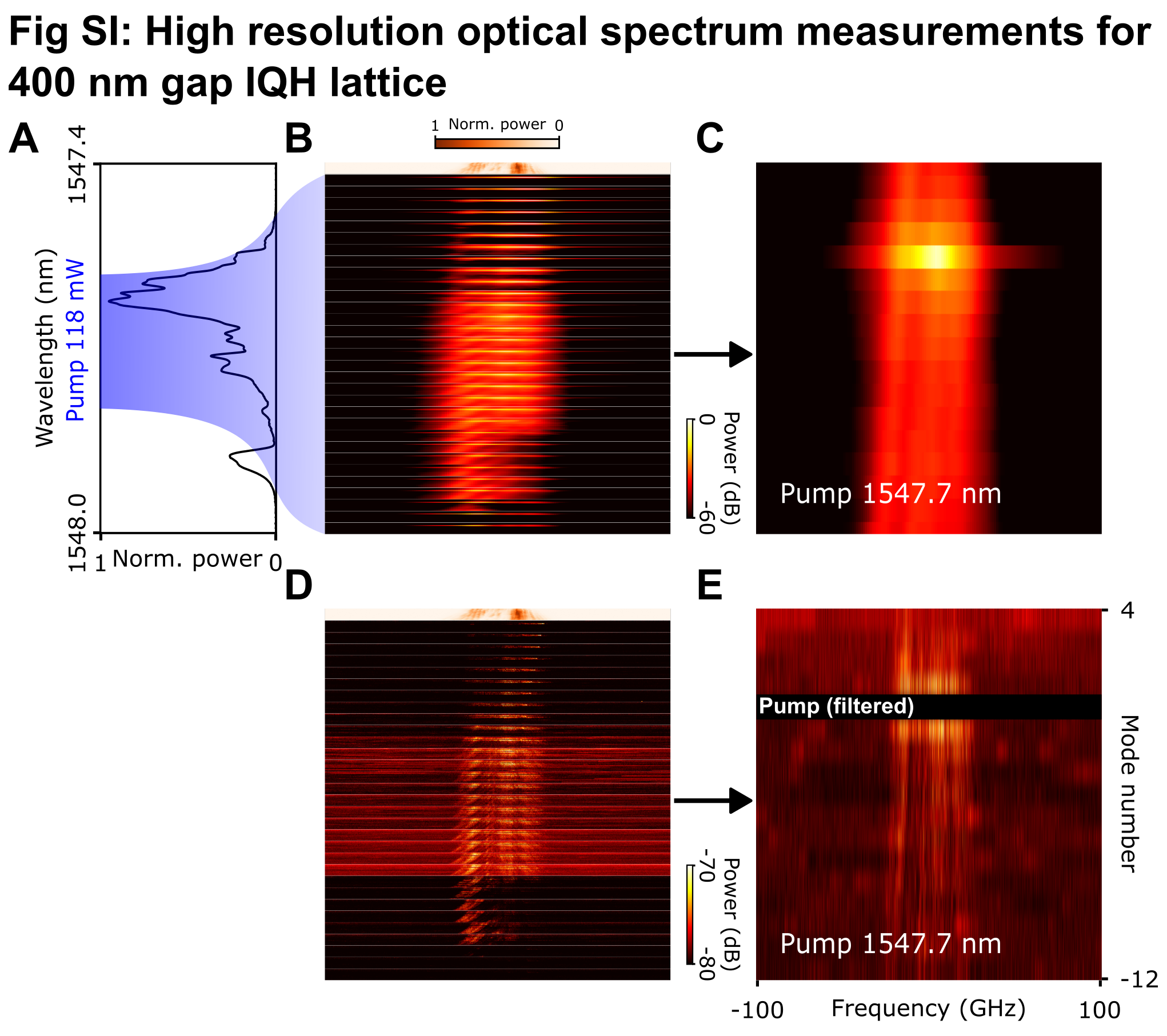}
\caption{\textbf{(A)} The linear drop spectrum of a 400 nm gap IQH device. We generate frequency combs by sweeping the pump wavelength across the topological bands with a fixed pump power of 118 mW, as indicated in blue shades. \textbf{(B)} Orange: the linear integrated dispersion plot of the device. Red: the integrated dispersion of the generated IQH combs measured with a 20 pm low-resolution optical spectrum analyzer. \textbf{(C)} Zoomed-in plot of the comb for pump wavelength 1547.70 nm. \textbf{(D-E)} Same plots as panels B-C for the same lattice, where the combs are measured with a 40 fm high-resolution optical spectrum analyzer.}
\label{FigSI:apex_400nm_sweep}
\end{figure*}
\newpage
%
%
\section{Topological IQH solitons over multi-modal AGFs}\label{sm:LLE_soliton}
In this section, we numerically solve the LLE combined with the dispersive tight-binding model defined over multi-modal AGFs, using the formalism described in the SI section~\ref{sm:lle}:
\begin{equation}
\begin{aligned}
\frac{d a_{m,\mu}}{d t}
&= i\delta\,a_{m,\mu}
   \;+i\mathcal{H}_{m,\mu}a_{m,\mu} +\, i\,\mathcal{FT}\mathbf{\{}
     \bigl|E_{m,\theta}\bigr|^{2} E_{m,\theta}\mathbf{\}}\,  \\
&\quad -\,\bigl(\kappa_{\mathrm{ex,\mu}}\,\delta_{m,\mathrm{IO}} + \kappa_{\mathrm{in}}\bigr)\,a_{m,\mu}
     \;+\; \delta_{m,\mathrm{IO}}\delta_{\mu,0}\,\mathcal{F}\,.
\end{aligned}
\end{equation}

Within this framework, we numerically demonstrate the first IQH topological soliton obtained by pumping the edge band, as shown in Fig.~\ref{FigSI:LLE_SOLITON}. The resulting nonlinear state exhibits soliton formation in each individual edge ring, together with a collective edge super-soliton, forming a nested soliton structure. While nested solitons in topological photonic systems have been numerically reported previously~\cite{mittal2021topological,huang2024hyperbolic}, those studies are restricted to conventional narrow-band tight-binding models. By contrast, our simulations incorporate the full multi-mode, frequency-dependent lattice parameters, enabling us not only to faithfully capture the nesting behavior but also to reveal a slope change in the integrated dispersion plot, which plays a crucial role in stabilizing the multi-mode soliton. A video of the IQH topological soliton is shown in the supplementary movie M1.

Fig.~\ref{FigSI:LLE_defect} presents a second example of IQH topological solitons in the presence of a lattice defect, where a ring is removed from the lattice boundary, and the system is pumped within the clockwise (CW) edge band. In this case, we observe the coexistence of two solitons forming a topological soliton molecule. Remarkably, the solitons circumvent the defect by dynamically reconstructing a new effective edge and continue to propagate unidirectionally along the CW direction. These observations provide direct evidence that the solitons inherit the topological protection of the underlying IQH edge states, demonstrating robustness against structural defects beyond what is expected for conventional Kerr solitons inside single rings. A video of the IQH topological soliton in the presence of the defect is shown in the supplementary movie M2.
While only two exemplary simulations of IQH topological solitons have been demonstrated here, we note that, going forward, one interesting path is the multi-dimensional phase diagram and stability analysis of these IQH topological solitons in the presence of the corrected gauge parameters $D_\mathrm{int},J(\mu),\phi(\mu)$.  

For simulations performed in Figs.~\ref{FigSI:LLE_SOLITON} and~\ref{FigSI:LLE_defect}, we choose the dispersive tight-binding Hamiltonian that corresponds to the lattice with a gap of 300 nm. The extrinsic loss rate $\kappa_{\rm{ex}}$ represents the evanescent coupling between the input/output ring and the coupling waveguides. Therefore, it is obtained from 3D FDTD simulations of directional couplers with realistic geometries. The intrinsic loss rate $\kappa_{\rm{in}}=0.35$~GHz is treated as a constant. The normalized pump field $\mathcal{F}=4.5$, which corresponds to an estimated power of 22~W. For each fixed detuning, we perform temporal evolution using the split-step and fourth-order Runge-Kutta methods. The number of temporal iterations for each detuning is set to 24,000, with the step size being 0.1/$J_0$, where $J_0$ is the coupling strength for the mode $\mu=0$. The detuning sweep is performed from high to low detunings, with a sweep step of $-0.0005~J_0$. 
We perform these simulations on workstations equipped with 128
GB of RAM, 3.2 GHz CPU cores, and multiple terabytes of hard drive storage. Consequently, each simulation presented in Figs.~\ref{FigSI:LLE_SOLITON} and~\ref{FigSI:LLE_defect} required 24 hours to complete and 1 TB of hardware space.

\begin{figure*}[h]
\centering
\includegraphics[width=1\textwidth]{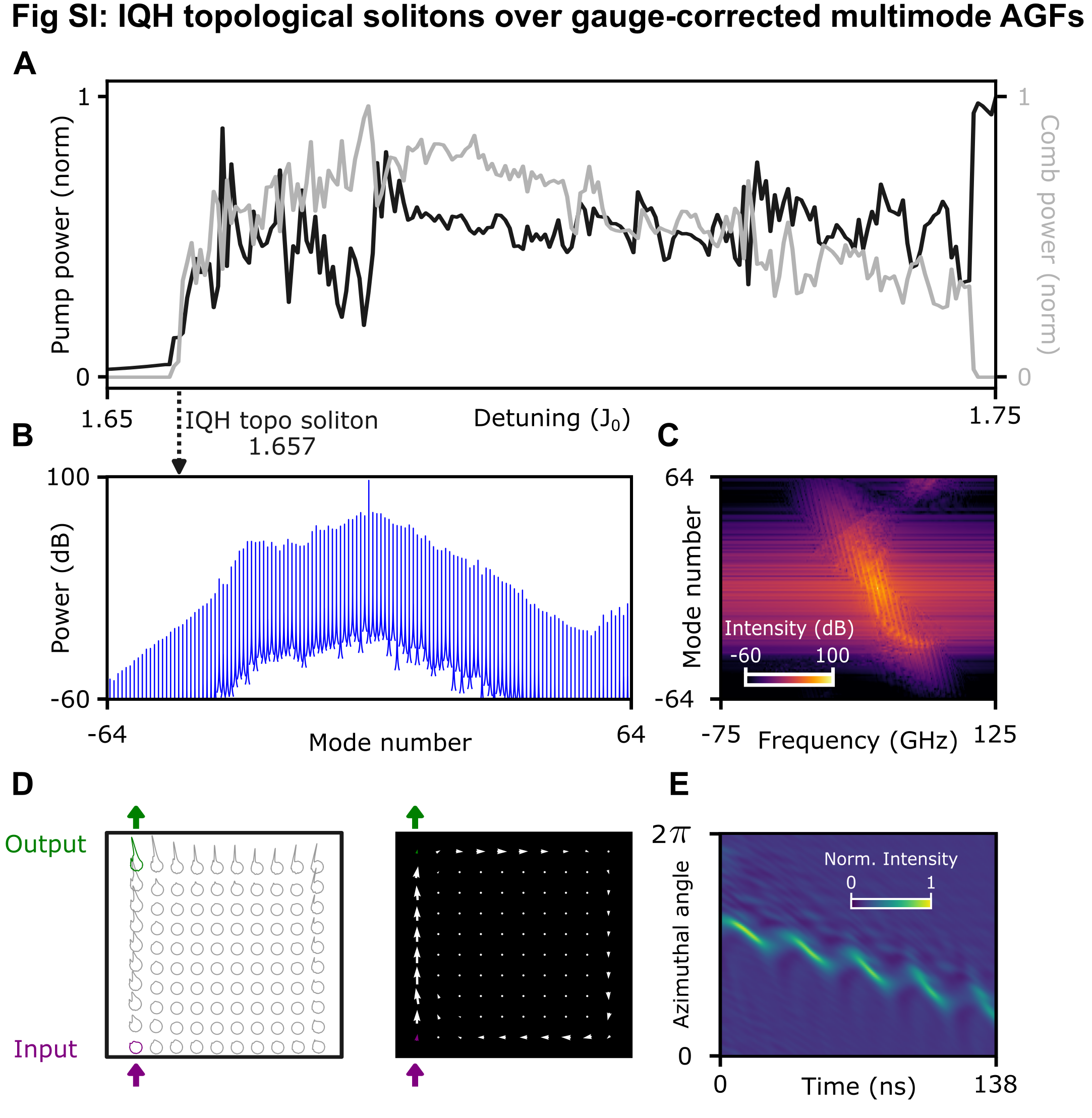}
\caption{\textbf{(A)} The pump mode power and the comb power as a function of pump detuning inside the output ring. $J_0$ is the coupling strength for mode $\mu=0$. \textbf{(B)} The IQH topological comb spectrum generated by pumping with the detuning 1.657~$J_0$. \textbf{(C)} The integrated dispersion plot of the topological IQH comb in (B). \textbf{(D)} Left: a temporal snapshot of the power profile for the IQH soliton comb in 3D. Right: the directional flow of photons in the mode $\mu=1$, demonstrating clockwise transport along the edge. \textbf{(E)} Spatiotemporal dynamics inside the output ring, featuring periodic temporal patterns.}
\label{FigSI:LLE_SOLITON}
\end{figure*}
\begin{figure*}[h]
\centering
\includegraphics[width=1\textwidth]{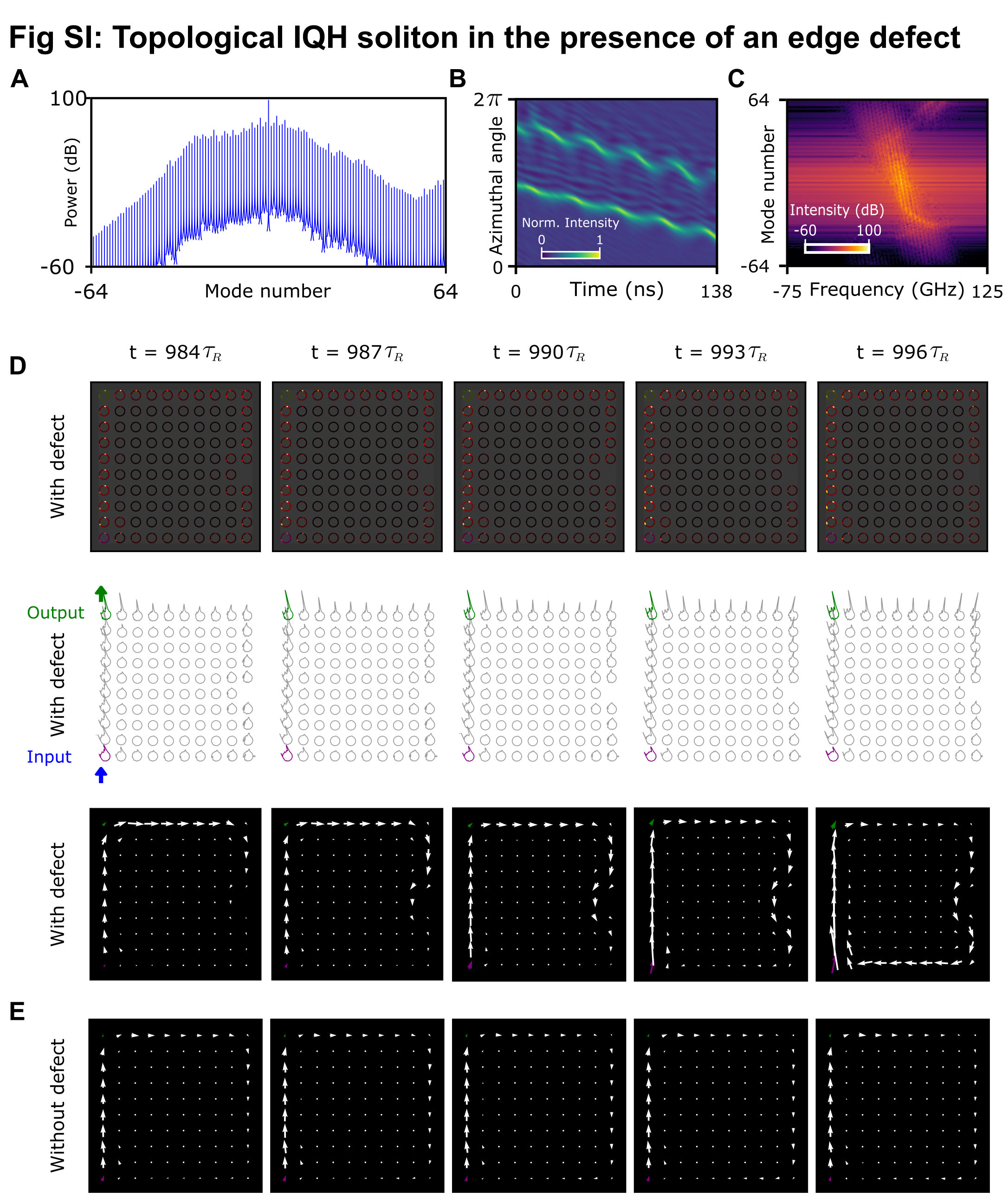}
\caption{\textbf{(A-C)} Exemplary comb spectrum \textbf{(A)}, the spatiotemporal plot \textbf{(B)}, and the integrated dispersion \textbf{(C)} of a two-IQH-soliton state in the presence of an edge defect. \textbf{(D)} Five temporal snapshots of the power profile in 2D (first row), 3D (second row), as well as the directional flow of photons in the mode $\mu=1$ (third row) for the lattice with an edge defect. These plots demonstrate topological protection against lattice defects in the nonlinear regime. \textbf{(E)} Five snapshots of the directional flow of photons in the mode $\mu=1$ for a lattice without defect. }
\label{FigSI:LLE_defect}
\end{figure*}
\newpage
\section{Wafer-scale reproducibility}\label{sm:wafer}
To demonstrate the robustness of the IQH comb and integrated dispersion profiles against fabrication variations, we performed comb generation experiments on devices with identical designs on three different chips fabricated in a multi-project wafer (MPW) facility, as shown in Figure \ref{FigSI: wafer scale}. For the lattices with gaps of 300 nm, where the evanescent coupling $J$ dominates the waveguide dispersion, even though we observed spectral shifts of up to $\approx$~180 GHz in the linear drop port transmission spectra, the measured linear drop spectrum, comb spectra, and integrated dispersion are consistent across all three devices with the same design. In contrast, when the lattice gaps are 400 nm, corresponding to a dominant waveguide dispersion over the evanescent coupling, we observed an absence of robustness in the linear drop spectrum. 
\begin{figure*}[h]
\centering
\includegraphics[width=1\textwidth]{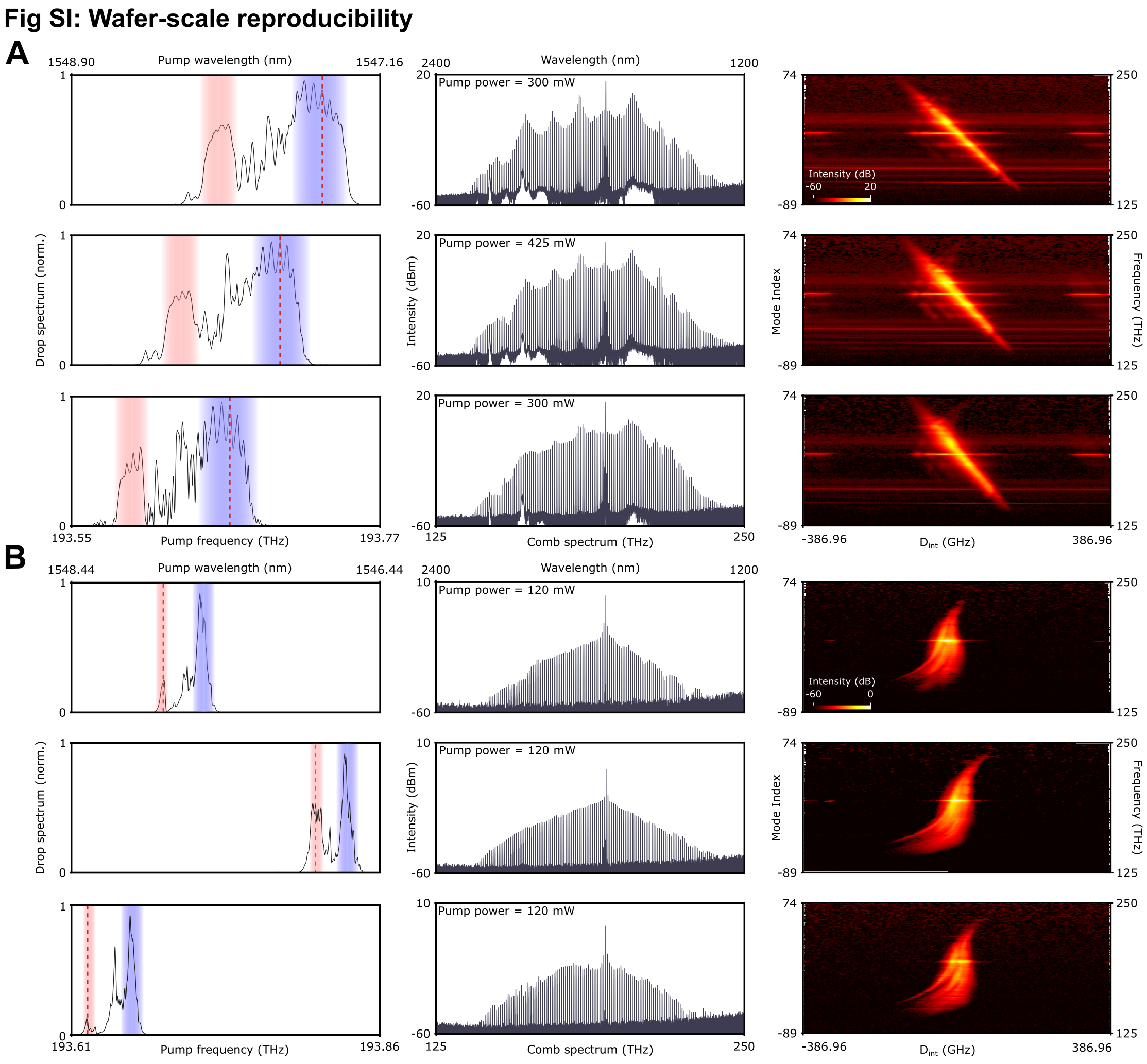}
\caption{Measured linear-regime drop port transmission spectra, comb spectra, and the corresponding integrated dispersion of devices with identical designs fabricated on three different chips on the same wafer. Two designs are measured where \textbf{(A)} J $\approx$ 20 GHz \textbf{(B)} J $\approx$ 10 GHz near 1548 nm. For each device, the pump wavelength is marked by red dashed lines in the drop spectrum. }
\label{FigSI: wafer scale}
\end{figure*}

\end{supplement}

\clearpage
\bibliography{Main.bib}
\end{document}